\documentstyle[12pt,epsf]{ioplppt2}
\begin{document}
\jl{1}
\bibliographystyle{prsty}

%\draft
%\preprint{}

\title{Finite-connectivity systems as error-correcting codes}
\author{Renato Vicente\footnote{vicenter@aston.ac.uk} and  David Saad\footnote{saadd@aston.ac.uk} }

\address{
The Neural Computing Research Group, Aston University, Birmingham B4 7ET,
UK}
\author {Yoshiyuki Kabashima\footnote{kaba@dis.titech.ac.jp}}
\address{ Department of Computational Intelligence and Systems Science,
Tokyo Institute of Technology, Yokohama 226, Japan}

\date{\today}

\begin{abstract}
We investigate the performance of parity check codes using the mapping
onto Ising spin systems proposed by Sourlas. We study codes where each parity
check comprises products of $K$ bits selected from the original digital
message with  exactly $C$ checks per  message bit.
We show, using the replica method,  that these codes saturate  
Shannon's coding bound for $K\rightarrow\infty$ when the
 code  rate $K/C$ is finite. 
We then examine the  finite temperature case  to asses the use
of simulated annealing methods for decoding, study  the performance
of the finite $K$ case and  extend the analysis to 
accommodate different types of noisy channels. The connection  between 
statistical physics and  belief propagation decoders is discussed and
the dynamics of the decoding itself is analyzed.  Further insight
into new approaches for improving the code performance is given.
\end{abstract}

%\maketitle
%\narrowtext

%@@@@@@@@@@@@@@@@@@@@@@@@@@@@@@@@@@@@@@@@@@@@@@@@@@@@@@@@@@@@@@@@@@@@@@

\section{Introduction}
\label{sec:intro}
Error-correction is  required whenever information has to be reliably
 transmitted through a noisy environment. The theoretical
 grounds for classical error-correcting codes were first presented in 
1948 by Shannon \cite{shannon}. He showed that it is possible to transmit 
 information trough a noisy channel with a vanishing error probability
 by encoding 
 up to a  given critical rate $R_c$ equivalent to the 
{\it channel capacity}.
 However, Shannon's arguments were non-constructive and devising such
 codes turned out to be a major practical problem in  the area of information 
transmission.  

In 1989  Sourlas \cite{sourlas89,sourlas94} proposed that, due to the
equivalence between addition over the field $\{0,1\}$ and multiplication over
$\{{\pm 1}\}$,  many error-correcting codes can be mapped  onto many-body 
spin-glasses with appropriately defined couplings. This observation opened
the possibility of applying techniques from statistical physics to 
study coding systems, in particular, these ideas were applied to the study of 
parity check codes.
These linear block codes can be represented by  matrices of  $N$ columns
and $M$ rows  that transform  $N$-bit messages to  $M$ ($>N$) parity checks.
Each  row represents bits involved in a 
particular check and each column represents checks  involving
the particular bit. The number of bits used in each check and the number 
of checks per  bit depends on the code construction. We concentrate on the 
case where  exactly $C$ checks are performed for each bit and exactly 
$K$ bits compose each check. 

The  {\it code rate} $R$ is  defined as the information conveyed per channel 
use $R=H_2(f_s)N/M=H_2(f_s)K/C$, where $H_2(f_s)=
-(1-f_s)\;\mbox {log}_2 (1-f_s)\;-
\;f_s\;\mbox {log}_2 (f_s) $ is the binary entropy  of the message with bias 
$f_s$. 

% FIGURE 1
% Encoding and Decoding
\begin{figure}
\hspace*{.4cm}
\epsfxsize=150mm  \epsfbox{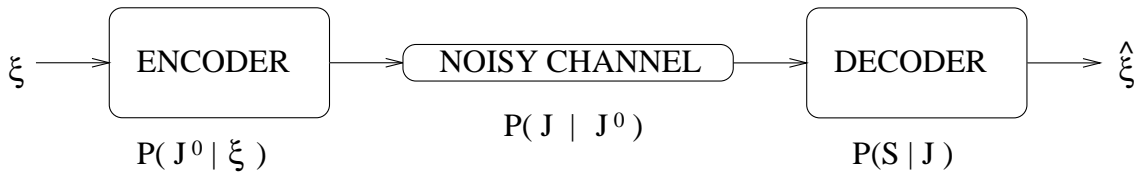}
\vspace{0.5cm}
\caption{The encoding, message  corruption in the noisy channel and 
 decoding can be represented  as  a Markovian process. The aim is to obtain 
a good estimative $\mbox{\boldmath $\widehat {\xi}$}$ for the  
original message $\xi$.}
\label{encode}
\end{figure}

In the mapping proposed by Sourlas a message is represented by a  
binary vector $\mbox{\boldmath $\xi$}
\in\{\pm 1\}^N$  encoded to a higher dimensional  vector
$\mbox{\boldmath $J^0$}\in\{\pm 1\}^M$ defined as $J^{0}_{\langle i_{1},
i_{2} \ldots i_{K}\rangle} = \xi_{i_{1}} \xi_{i_{2}} \ldots \xi_{i_{K}}$,
where $M$ sets of $K$ indices are randomly  chosen. A corrupted version 
$\mbox{\boldmath $J$}$ of the encoded message $\mbox{\boldmath $J^0$}$ has to 
be decoded for retrieving the original message.  The decoding process can be 
viewed as a statistical Bayesian process \cite{iba98} (see Fig.\ref{encode}). 
 Decoding focuses on  producing an estimate 
 $\widehat{\mbox{\boldmath $\xi$}}$ to  the original  message  that minimizes a given expected 
loss $\langle\langle {\cal L}(\xi,\widehat{\xi})\rangle_{p(J\mid\xi)}\rangle_
{p(\xi)}$ averaged  over the indicated probability distributions. The 
definition of the loss depends on the particular task; the simple Hamming 
distance ${\cal L}(\xi,\widehat{\xi})=\sum_j \xi_j \widehat{\xi}_j$ can be 
used for decoding binary messages. An optimal estimator for this particular
loss function is  $\widehat{\xi}_j=\mbox{sign}\langle S_j 
\rangle_{p(S\mid J)}$ \cite{iba98}, where $\mbox{\boldmath $S$}$ 
is a $N$ dimensional binary vector  representing outcomes of the 
decoding process.
 Using Bayes' theorem, the posterior 
probability can be written as $\mbox{ln }p(\mbox{\boldmath $S$}\mid\mbox{\boldmath $J$} )=\mbox{ln }p(\mbox{\boldmath $J$}\mid \mbox{\boldmath $S$})
+ \mbox{ln }p(\mbox{\boldmath $S$}) + \mbox{const}$. 
Sourlas has shown  \cite{sourlas94}  that for  parity check codes this 
posterior  can be written as  a many-body Hamiltonian:
\begin{eqnarray}
\label{eq:Hamiltonian} 
\mbox{ln }p(\mbox{\boldmath $S$}\mid\mbox{\boldmath $J$} )&=&-\beta\; {\cal H}(\mbox{\boldmath $S$})\nonumber\\
&=&\beta \sum_{\mu} 
{\cal A}_{\mu} \ J_{\mu} \ \prod_{i\in\mu} S_{i} + 
\beta{\cal H}_{\mbox{\scriptsize  prior}} (\mbox{\boldmath $S$}),
\end{eqnarray}
where $\mu=\left\langle i_{1},\ldots i_{K} \right\rangle$ is a set of indices 
and ${\cal A}$ is a  tensor with the properties  ${\cal A}_\mu\in\{0,1\}$ and 
 $\sum_{\mu\setminus i}{\cal A}_\mu=C$ $\forall i$, which determines the $M$ 
components of the codeword $\mbox{\boldmath $J$}^{0}$. The second term ${\cal H}_{\mbox{\scriptsize  prior}} (\mbox{\boldmath $S$})$ stands for 
the  prior knowledge on  the actual  messages; it can
be chosen as ${\cal H}_{\mbox{\scriptsize  prior}}(\mbox{\boldmath $S$})=F\sum_{j=1}^{N} S_j$
 to represent the expected bias in the 
message bits. For the simple case of a memoryless binary symmetric channel
(BSC), $\mbox{\boldmath $J$}$   is a corrupted version of the
 transmitted message  $\mbox{\boldmath $J$}^{0}$ where each bit is 
independently  flipped  with probability $p$ during 
 transmission. The  hyper-parameter $\beta$, that reaches 
an  optimal value at  Nishimori's temperature \cite{iba98,rujan93,nishi},
 is related to  the channel corruption rate.
The decoding procedure translates to finding the thermodynamical spin 
averages for the system defined by the Hamiltonian (\ref{eq:Hamiltonian}) at 
a certain temperature (Nishimori's temperature for optimal decoding); as the 
original message is binary, the retrieved message bits are given by the signs 
of the corresponding averages.

In the statistical physics framework the  performance  of the  
error-correcting process can be measured by the overlap between actual 
message and estimate for a given scenario characterized by a code rate, 
corruption process and information content of the message. To asses
the typical properties we average this overlap  over all possible
codes $\cal A$ and  noise realizations (possible corrupted vectors
$\mbox{\boldmath  $J$}$) given the message $\mbox{\boldmath $\xi$}$ and
then over all possible messages: 

\begin{equation}
\label{eq:mag}
m=\frac {1}{N}\left \langle  \sum_{i=1}^N {\xi}_i \;\left \langle 
\mbox{sign}\langle S_i \rangle \right\rangle_{{\cal A},J|\xi}\right
\rangle_{\xi}
\end{equation}
Here $\mbox{sign}\langle S_i \rangle$  is the sign of the spins thermal 
average corresponding to the Bayesian optimal decoding. The average error 
per bit  is then given by $p_e = (1-m)/2 $. Although this performance measure is not the usual physical magnetization (it can be  better described as a measure of misalignment of the decoded message), for brevity, 
 we will refer  to it as {\it magnetization}.

From the statistical physics point of view, the number of checks per bit is 
analogous to the  spin system connectivity  and the number of
bits in each check is analogous to the number of spins per interaction.
Sourlas' code  has been studied in the case of extensive connectivity 
, where   the
number of bonds $C \!\sim\!$ \scriptsize $\left( \begin{array}{c} N-1 \\ K-1
\end{array} \right)$ \normalsize scales with the system size. In
this case  it can be  mapped  onto known problems
in statistical physics such  as the SK \cite{SK} ($K\!\!=\!\!2$) and 
Random Energy  (REM)  \cite{Derrida_REM} ($K \!\!\rightarrow\!\! 
\infty$) models. It has been shown   that the  REM saturates  
Shannon's bound \cite{sourlas89}.  However, it has a rather limited practical 
relevance  as the choice of extensive connectivity corresponds to a  
vanishingly small code rate.

Here we  present an analysis of   Sourlas' code for the case of finite
connectivity where the code rate is non-vanishing,
detailing and extending our previous brief reports  \cite{ks98a,ks98b}.
We show that  Shannon's bound can also be attained at  finite code rates.
We study the decoding dynamics  and discuss the connections between 
statistical physics and belief propagation methods.

 This paper is organized as follows: in Section II  we introduce a naive 
mean-field
model that contains all the necessary ingredients to understand the system qualitatively. Section III  describes  the statistical
 physics treatment of Sourlas' code showing that  Shannon's bound
 can  be attained for finite code  rates if $K\rightarrow\infty$.  The finite $K$ case and the Gaussian noise are also discussed in Section III.  The decoding dynamics is  analyzed in Section IV. Concluding remarks are given in 
Section V.  Appendices with detailed calculations are also provided.

%@@@@@@@@@@@@@@@@@@@@@@@@@@@@@@@@@@@@@@@@@@@@@@@@@@@@@@@@@@@@@@@@@@@@@@@@@

\section{Naive Mean Field Theory}
\label{sec:naive} 

\subsection {Equilibrium}

% FIGURE 2
% Naive model
\begin{figure}
\hspace*{.4cm}
\epsfxsize=120mm  \epsfbox{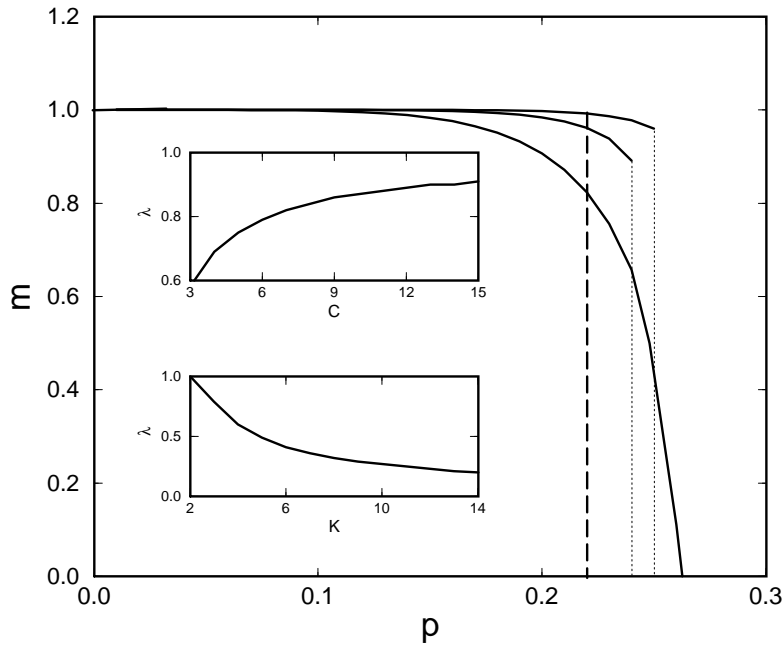}
%\vspace{0.5cm}
\caption{Code performance measured by the magnetization $m$ as a function 
of the noise level $p$ as  given by the naive mean-field theory at code rate 
$R=1/2$ and $K=2,3,4$ respectively from the bottom. The long-dashed line 
indicates  PARA-FERRO coexistence. Insets: Maximum initial 
deviation $\lambda$ for convergence at a noise level $p=0.1$. Top inset:
 $K=3$ and increasing $C$. Bottom inset: Code rate  $R=1/2$ and 
increasing $K$.}
\label{naive}
\end{figure}

To gain some  insight into the code behavior one can start by considering 
that the original message is $\xi_j=1$ for all $j$ (so $m=1$ will correspond 
to perfect decoding) and   use  Weiss' mean-field theory as a first (naive) 
approximation. The idea is to consider an
effective field given by (for unbiased messages with $F=0$):
\begin {equation}
\label{eq:naiveeff}
h^{\mbox{\scriptsize eff}}_j =\sum_{\{\mu:j\in \mu\}}J_\mu \prod_{i\in\mu\setminus j} S_i
\end{equation}
acting in every site. The first strong approximation here consists in
 disregarding 
the reaction fields that describe the influence of site $j$ back over the 
system.
The local magnetization can then be calculated:
\begin {equation}
\label{eq:naivemag}
m_j=\left\langle\mbox{tanh}\left(\beta h^{\mbox{\scriptsize eff}}_j \right)\right\rangle_{J,S}\simeq \mbox{tanh}\beta \left\langle h^{\mbox{\scriptsize eff}}_j \right\rangle_{J,S},
\end{equation}
where we introduced a further approximation taking averages inside the 
function that can be seen as a high  temperature approximation. Disregarding correlations among spins and  computing 
the proper averages one can write:
\begin {equation}
\label{eq:naivemagav}
m=\mbox{tanh} \left( \beta \; C (1-2p)\; m^{K-1}\right),
\end{equation}
where $p$ is the noise level in the channel. An alternative way to derive the above equation is by considering the free-energy:
\begin {equation}
\label{eq:naivefree}
f(m)=-(1-2p)\frac{C}{K}m^{K}-\frac{s(m)}{\beta}.
\end{equation}
The entropic  term $s(m)$ is:
\begin {equation} 
s(m)=-\frac{1+m}{2}\mbox{ ln}
\left(\frac{1+m}{2}\right)-\frac{1-m}{2}\mbox{ ln}
\left(\frac{1-m}{2}\right).
\end{equation}
Minimizing this free-energy one can obtain Eq.(\ref{eq:naivemagav}) whose
solutions give the possible phases  after the decoding process.  In Fig.
\ref{naive}  we show the maximum magnetization solutions $m$ for 
Eq.(\ref{eq:naivemagav}) as a function of the flip rate $p$ at code rate 
$R=1/2$ and $K=2,3,4$. For $K=2$ the performance degrades faster with the 
noise level than  in the $K>2$ case. The dashed line indicates coexistence 
between paramagnetic  (PARA) $m=0$ and ferromagnetic (FERRO) $m>0$ phases.

\subsection{Decoding Dynamics}

% FIGURE 3
% Node
\begin{figure}
\hspace*{3cm}
\epsfxsize=50mm  \epsfbox{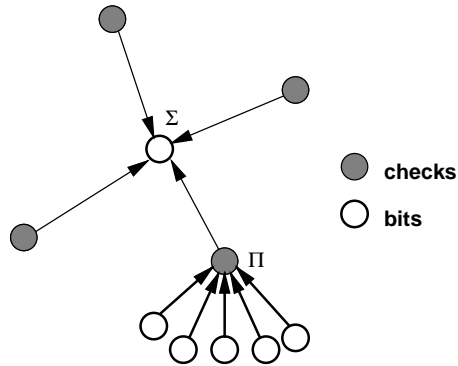}
\vspace{0.5cm}
\caption{Graph representing a  code.}
\label{node}
\end{figure}

In a naive mean-field framework the decoding process can be 
seen as an iterative solution for (\ref{eq:naivemagav}) starting from a
magnetization value that depends on the prior knowledge about the 
original message. The fixed points of this dynamics correspond to the minima of the free-energy; a specific minimum is reached depending on the initial condition. In the insets of Fig.\ref{naive} we show,  as a measure for the basin of attraction, the maximal deviation between
the initial condition and the original message  $\lambda=1-m_0$  that allows  
convergence to a FERRO solution. At the bottom inset we show the deviation 
$\lambda$ at code rate $R=1/2$, increasing  values of $K$ and 
 noise level $p=0.1$ . 
An increasing initial magnetization is needed when $K$ increases, 
decoding without prior knowledge is only possible for $K=2$. The top 
inset shows $\lambda$ for $K=3$, $p=0.1$; as $C$ increases (code rate decreases), the basin of attraction increases. 

One can understand intuitively how the basin of attraction depends on 
the  connectivities by representing the code in a graph with bit and check nodes and looking at the mean-field  
behavior of a  single bit node (see Fig.\ref{node}). The corrupted checks
 contribute  wrong ($-1$ for the ``all ones'' message case) values to the  bit nodes ($m<1$ in the mean field). Since  check node values correspond to a product of $K-1$ bit values, the probability of  updating these nodes to the wrong values increases with $K$, degrading
the overall performance.  On the other hand, if $C$ increases   for a fixed $K$ the bit nodes  gather more information and are  less sensitive to  the presence of (a limited amount of ) wrong bits .

Although this naive picture indicates some of the qualitative features 
of real codes, one certainly cannot rely in its numerical predictions. 
In the following sections we will study Sourlas' codes using more 
sophisticated  techniques that will substantially refine the analysis.

%@@@@@@@@@@@@@@@@@@@@@@@@@@@@@@@@@@@@@@@@@@@@@@@@@@@@@@@@@@@@@@@@@@@@@@@@@
\section{Equilibrium}

%@@@@@@@@@@@@@@@@@@@@@@@@@@@
\subsection{Replica Theory} 
\label{sec:replica}

In the following subsections we will develop the replica symmetric theory for Sourlas' codes and show that, in addition to providing a good description of the equilibrium, it describes the typical  decoding dynamics using  
 belief propagation methods. 

The previous naive ``all ones'' messages assumption can be formally translated
to  the  gauge transformation \cite{frad} $S_{i} \!\!\mapsto\!\!
 S_{i} \xi_{i}$ and $J_{\mu}\!\!\mapsto\!\! J_{\mu}\prod_{i\in \mu} \xi_{i} $ 
that maps  any general  message to the FERRO configuration defined as 
$\xi_i^{*}=1$  $\forall i$. One can then rewrite  the Hamiltonian in the form:
\begin{equation}
\label{eq:Hamiltonian_gauge} {\cal H}(\mbox{\boldmath $S$})=- 
\sum_{\mu}  {\cal A}_{\mu} \ J_{\mu} \ \prod_{i\in\mu} S_{i} -
F \sum_{k}\xi_k  S_{k} \ ,
\end{equation} 

With this transformation, the bits of the uncorrupted encoded message are 
 $J^0_i=1$ $\forall i$ and, for a BSC,  the corrupted bits are random variables with probability:
\begin{equation}
\label{eq:xi_J_prob_dist}
{\cal P}\left(J_{\mu}\right) =  (1\!-\!p)\ \delta \left(J_{\mu} \!-\! 1 \right) +  p \ \delta\left(J_{\mu} \!+\! 1 \right), 
\end{equation}
where $p$ is the channel flip rate. For deriving typical properties of these codes one has  obtain an expression for the free-energy by invoking the replica approach where the free-energy is defined as:
\begin{equation}
\label{eq:freenergy}
f= -\frac{1}{\beta}\lim_{N\rightarrow\infty} \frac{1}{N}
\left.\frac{\partial} {\partial {\mathit n}}\right |_{{\mathit n}=0} \langle {\cal Z}^{ \mathit n}\rangle_{{\cal A},\xi,J}, 
\end{equation}
where $\langle {\cal Z}^{ \mathit n}\rangle_{{\cal A},\xi,J}$ represents
an analytical  continuation in the interval  $n\in[0,1]$ of  the replicated
partition function defined as:

\begin{equation}
\label{eq:partit}
\langle {\cal Z}^{n}\rangle_{{\cal A},\xi,J} = \mbox{Tr}_{\{S_j^\alpha\}}
\left[\left \langle e^{ \beta F \sum_{\alpha,k}\xi_k  S^\alpha_{k}}\right 
\rangle_{\xi}\left\langle \exp\left(\beta\sum_{\alpha,\mu} 
{\cal A}_{\mu} \ J_{\mu} \ \prod_{i\in\mu} S^\alpha_{i} \right) 
\right\rangle_{{\cal A},J} \right]. 
\end{equation}

 The  magnetization can be 
rewritten in the gauged variables as :
\begin{equation}
\label{eq:mag_gauged}
m= \left \langle\left \langle  \mbox{sign}\langle S_i \rangle 
\right\rangle_{{\cal A},J|\xi^*}\right \rangle_{\xi},
\end{equation}
where $\xi^*$  denotes the transformation of a message $\xi$ into 
the FERRO configuration. The usual magnetization per site can be  
easily obtained by calculating  
\begin{equation}
\label{eq:fundamental}
\left \langle\left \langle S_i \right \rangle\right\rangle_{{\cal A},J,\xi}=-  \left( \frac {\partial f} { \partial(\xi F)} \right). 
\end{equation}
From this derivative one can find the distribution of the effective  local 
fields $h_j$ that can be used to asses the  magnetization $m$, since 
$\mbox{sign}\left(\langle S_j\rangle\right)=\mbox{sign}(h_j)$ .

To compute the replicated partition function  we  closely follow 
Ref. \cite{wong_a}. We  average uniformly over all codes  ${\cal A}$ such 
that $\sum_{\mu\setminus i}{\cal A}_{\mu}= C$ $\forall i$ to  find: 

\begin{eqnarray}
\label{eq:partit_2}
\langle {\cal Z}^{n}\rangle_{{\cal A},\xi,J}& =&\exp \left\{ N \;Extr_
{q,\widehat{q}}\left[C-\frac{C}{K}+\frac{C}{K}\left(\sum_{l=0}^{n}{\cal T}_l\sum_{\langle \alpha_1 \ldots \alpha_l\rangle}
q_{\alpha_1 \ldots \alpha_l}^{K}  \right)\right.\right.\nonumber\\ 
& -&\left.\left. C \left(\sum_{l=0}^{n}\sum_{\langle \alpha_1 \ldots
\alpha_l\rangle}q_{\alpha_1 \ldots \alpha_l}\widehat{q}
_{\alpha_1 \ldots \alpha_l}\right) \nonumber \right. \right.\\ 
&+& \left.\left.\ln \mbox{Tr}_{\{S^{\alpha}\}}\left \langle e^{\beta  F\xi\sum
_{\alpha}S^\alpha}\right\rangle_{\xi}\left(\sum_{l=0}^{n} 
\sum_{\langle \alpha_1 \ldots \alpha_l\rangle}\widehat{q}
_{\alpha_1 \ldots \alpha_l}S^{\alpha_1}\ldots S^{\alpha_l} \right)^C \right]\right\}, 
\end{eqnarray}
where ${\cal T}_l=\langle  \tanh^l(\beta J) \rangle_J$, as 
in  \cite{viana}, and $q_0=1$. We give details of this calculation in 
the Appendix A.
At the extremum the order parameters acquire expressions similar to those of
 Ref. \cite{wong_a}:

\begin{eqnarray}
\label{order-param}
\widehat{q}_{\alpha_1,...,\alpha_l}&=& {\cal T}_l\;  q^{K-1}_
{\alpha_1,...,\alpha_l}\nonumber \\
q_{\alpha_1,...,\alpha_l}&=&\left \langle \left (\prod_{i=1}^l
S^{\alpha_i} \right) \left(\sum_{l=0}^{n}\sum_{\langle \alpha_1 \ldots 
\alpha_l\rangle}\widehat{q}_{\alpha_1 \ldots \alpha_l}S^
{\alpha_1}\ldots S^{\alpha_l}\right)^{-1}\right\rangle_{\cal X}.
\end{eqnarray}
where 
\begin{equation}
{\cal X}=\left \langle e^{\beta  F\xi\sum_{\alpha}S^\alpha}\right\rangle_
{\xi}\left(\sum_{l=0}^{n}\sum_{\langle
\alpha_1 \ldots\alpha_l\rangle}\widehat{q}_{\alpha_1 \ldots \alpha_l}
S^{\alpha_1}\ldots  S^{\alpha_l} \right)^{C},
\end{equation}
and $\langle...\rangle_{\cal X}=\mbox{Tr}_{\{S^{\alpha}\}}\left[(...)
{\cal X}\right]/\mbox{Tr}_{\{S^{\alpha}\}}\left[(...)\right]$.
The term 
$\widehat{p}(\underline {S})=\sum_{l=0}^{n}\sum_{\langle \alpha_1 \ldots\alpha_l\rangle}
\widehat{q}_{\alpha_1 \ldots \alpha_l} S^{\alpha_1}\ldots  S^{\alpha_l}$ 
represents a probability distribution  
over  the space of replicas and $p_0(\underline{S})=\left \langle e^{\beta  F\xi\sum_{\alpha}S^\alpha}\right\rangle_{\xi}$ is a prior distribution over the same space. For reasons that will become clear in 
Section \ref{sec:decoding}, $q_{\alpha_1,...,\alpha_l}$ represents one
 $l$-th momentum of the
equilibrium distribution of a bit-check edge  in  a belief network during the
 decoding process and $\widehat{q}_{\alpha_1 \ldots \alpha_l}$ 
represents $l$-th moments of a check-bit edge
 equilibrium distribution . The distribution ${\cal X}$ represents the probability of a certain site (bit node) configuration  subjected to exactly $C$
interactions and with prior probability given by $p_0$.

%@@@@@@@@@@@@@@@@@@@@@@@@@@@@@@

\subsection{Replica Symmetric Solution}
\label{sec:symmetric}

The replica symmetric (RS) ansatz can be introduced via the auxiliary
fields $\pi(x)$ and $\widehat{\pi}(y)$ in the following way 
(see also \cite{wong_a}):

\begin{eqnarray}
\label{eq:auxfields}
\widehat{q}_{\alpha_1 ... \alpha_l}&=&\int \: dy \; \widehat{\pi}(y)
\tanh^l(\beta y) ,\nonumber\\
q_{\alpha_1 ... \alpha_l}&=&\int \: dx \; \pi(x) \tanh^l(\beta x)
\end{eqnarray}
for $l=1,2,\ldots$. 

Plugging it into the replicated partition function (\ref{eq:partit_2}), 
performing the limit $n\rightarrow 0$ and using Eq.(\ref{eq:freenergy}) 
(see Appendix \ref{app:B} for details) one obtains:

\begin{eqnarray}
\label{eq:freesym}
f&=&-\frac{1}{\beta}\: Extr_{\pi,\widehat{\pi}}\left \{\alpha \ln \cosh
\beta \right. \\
&+& \alpha \int \left[
\prod_{l=1}^{K} dx_{l} \ \pi(x_{l}) \right] \left\langle \ln \left[ 1
+ \tanh \beta J \ \prod_{j=1}^{K} \tanh \beta x_{j} \right]
\right\rangle_{J} \nonumber \\ 
&-& C \int dx \ dy \ \pi(x) \
\widehat{\pi}(y) \ \ln \left[ 1 + \tanh \beta x \ \tanh \beta y
\right] \nonumber\\
&-& C \int dy \ \widehat{\pi}(y) \ \ln \cosh \beta y \nonumber\\
 &+&\left.  \int \left[ \prod_{l=1}^{C} dy_{l} \ \widehat{\pi}(y_{l})
\right] \left\langle \ln \left[ 2 \cosh \beta \left(\sum_{j=1}^{C}
y_{j} + F \xi \right) \right]\right\rangle_{\xi} \right \}\nonumber,
\end{eqnarray}
where $\alpha=C/K$.
The saddle-point equations, obtained by varying
Eq.(\ref{eq:freesym}) with respect to the probability distributions,
provide a set of relations between $\pi(x)$ and $\widehat{\pi}(y)$
\begin{eqnarray}
\label{eq:saddle_point}
\fl\pi(x) &=& \int \left[ \prod_{l=1}^{C-1} dy_{l} \ \widehat{\pi}(y_{l})
\right] \ \left\langle \delta \left[ x - \sum_{j=1}^{C-1} y_{j} - F \xi
\right]\right\rangle_{\xi} \\ 
\fl \widehat{\pi}(y) &=& \int \left[
\prod_{l=1}^{K-1} dx_{l} \ \pi(x_{l}) \right] \ \left\langle \delta 
\left[ y -
\frac{1}{\beta} \tanh^{-1} \left( \tanh\beta J \ \prod_{j=1}^{K-1}
\tanh \beta x_{j} \right) \right] \right\rangle_{J} \ .  \nonumber
\end{eqnarray}

Later we will show that this self-consistent pair of equations can be seen as 
a mean-field version for  the belief propagation decoding.
Using Eq.(\ref{eq:fundamental}) one  finds that the local field distribution is
 :
\begin{equation}
\label{eq:local_field}
 P(h)=\int \left[ \prod_{l=1}^{C} dy_{l} \
\widehat{\pi}(y_{l}) \right] \ \left\langle \delta \left[ h -
\sum_{j=1}^{C} y_{j} - F \xi \right]\right\rangle_{\xi},
\end{equation}
where $\widehat{\pi}(y)$ is given by the saddle point equations above.

The  magnetization  (\ref{eq:mag}) can then be calculated using:
\begin{equation}
\label{eq:mag_sym}
m = \int d h \ \mbox{sign} (h) \, P(h).
\end{equation}

The code performance can be assessed by assuming a particular prior  distribution for the message bits,  
solving  the  saddle-point equations (\ref{eq:saddle_point}) numerically and 
then computing  the  magnetization. 

Instabilities in the solution within the space of symmetric replicas can be
probed looking at  second derivatives of the functional  whose extremum
defines the free-energy (\ref{eq:freesym}). The  simplest necessary
condition for stability is having non-negative second functional derivatives
in relation to $\pi(x)$ (and $\widehat{\pi}(y)$) :
  
\begin{equation}
\label{eq:stability}
\fl \frac{1}{\beta} \int \left[\prod_{l=1}^{K-2} dx_{l} \ 
\pi(x_{l}) \right] \left\langle
 \ln \left[ 1+ \tanh \beta J \ \tanh^2 \beta x\prod_{j=1}^{K-2} \tanh
 \beta x_{j} \right]
\right\rangle_{J} \geq 0,
\end{equation}
for all $x$.  
The replica symmetric solution is  expected to be unstable for 
sufficiently  low temperatures (large $\beta$). For high temperatures
we can expand the above expression around small $\beta$ to find the
stability condition:
\begin{equation}
\label{eq:stabhigh}
\langle J\rangle_{J}  \langle x \rangle_{\pi}^{K-2}\geq 0 \; 
\end{equation}
We expect the average $\langle x \rangle_{\pi}=\int dx\, \pi(x)\, x$ 
to be zero in PARA phase and positive in  FERRO phase,
satisfying the stability condition. This result is still generally inconclusive, but provides some evidence  that can be examined numerically. In Section \ref{sec:bound} we will  test the stability of our solutions using 
  condition (\ref{eq:stability}).

In the next sections we restrict our study to the unbiased case ($F=0$),
 which is of  practical relevance,  since it is always possible to compress a biased message to an unbiased one.

%@@@@@@@@@@@@@@@@@@@@@@@@@@@@@@@@@@@@

\subsection {Case $K\rightarrow\infty$, $C=\alpha K$  }
\label{sec:Kinfty}
For this case one can obtain solutions to the
saddle-point equations for arbitrary temperatures. In the  first saddle-point 
equation (\ref{eq:saddle_point}) one can write:
\begin{equation}
\label{central_limit}
x=\sum_{l=1}^{C-1} y_l \approx (C-1)\langle y \rangle_{\widehat{\pi}} = (C-1)\int dy \; y\;
\widehat{\pi}(y). 
\end{equation}  
It means that if $\langle y \rangle_{\widehat{\pi}}=0$ (as it is the in 
PARA  and spin glass  (SG) phases) 
then $\pi(x)$ must be concentrated at $x=0$ implying that 
${\pi}(x)=\delta(x)$ and $\widehat{\pi}(y)=\delta(y)$ are 
the only possible  solutions. Moreover, Eq.(\ref{central_limit})
implies that  in FERRO phase one  can expect 
$x\approx{\cal O}(K)$.

Using Eq.(\ref{central_limit}) and the second saddle-point equation 
(\ref{eq:saddle_point}) one can find a self-consistent equation for the 
mean-field $\langle y \rangle_{\widehat{\pi}}$:
\begin{equation}
\label{mean_field}
\langle y \rangle_{\widehat{\pi}} = \left \langle\frac{1}{\beta}\, \mbox{tanh}^{-1}
\left[\mbox{tanh}(\beta J)\left(\mbox{tanh}(\beta (C-1)\langle y \rangle_{\widehat{\pi}})
\right)^{K-1}  \right]\right\rangle_J.
\end{equation}
For a BSC the above average is over  distribution 
(\ref{eq:xi_J_prob_dist}). Computing the average, using  $C=\alpha K$
and rescaling the temperature as $\beta = \tilde{\beta} (\mbox {ln}K)/K $, 
in the limit  $K\rightarrow\infty$ one obtains: 
\begin{equation}
\label{mean_field2}
\langle y \rangle_{\widehat{\pi}} = (1-2p)\left[\mbox{tanh}
(\tilde{\beta}\alpha\langle y\rangle_{\widehat{\pi}}
\mbox{ ln}(K))  \right]^{K},
\end{equation}
where $p$ is the channel flip probability. 
The mean-field $\langle y \rangle_{\widehat{\pi}} = 0 $ is always a solution 
to  this equation (either 
PARA or SG); at $\beta_c =\mbox{ln}(K) /( 2\alpha K(1-2p))$
an extra  non-trivial FERRO solution emerges with 
$\langle y \rangle_{\widehat{\pi}}=1-2p$.  As the connection with
the magnetization $m$ is given by Eq. (\ref{eq:local_field}) and 
Eq. (\ref{eq:mag_sym}); it is not difficult to see that it implies $m=1$ for 
FERRO solution. One remarkable
point is that the temperature were the FERRO
solution emerges is 
$\beta_c \sim {\cal O}(\mbox{ln}(K)/K)$; it means that in a simulated
annealing process  PARA-FERRO barriers emerge quite early  for large $K$ 
values implying metastability and, consequently, a very slow convergence. 
It seems to advocate the use of small 
 $K$ values  in  practical applications. This case is  analyzed  in  Section
 \ref{sec:finite}. For $\beta>\beta_c$ both PARA and FERRO solutions exist.

% FIGURE 4
% Finite temperature phase diagram 
\begin{figure}
\hspace*{2cm}
\epsfxsize=120mm  \epsfbox{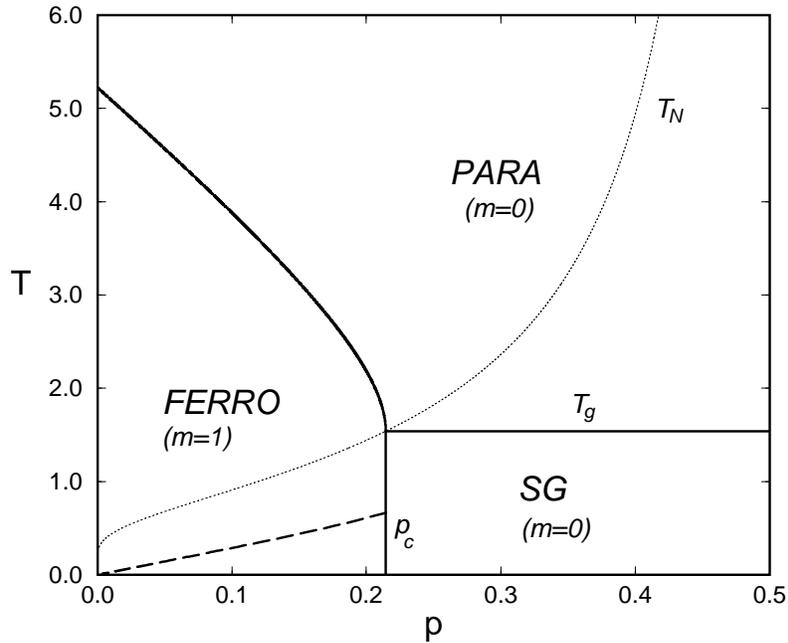}
\caption{Phase diagram in the plane temperature $T$ versus noise level $p$ 
for $K\rightarrow\infty$ and $C=\alpha K$, with
 $\alpha=4$. The dotted line indicates Nishimori's temperature $T_N$
. Full lines represent coexistence. The critical noise level is $p_c$.
 The necessary condition for stability in the FERRO phase is satisfied above 
the dashed line.}
\label{phase}
\end{figure}

The FERRO free-energy can be obtained from Eq.(\ref{eq:freesym}) using  Eq.(\ref{central_limit}), being $f_{\mbox{\scriptsize FERRO}}=-\alpha
(1-2p)$. The corresponding entropy is $s_{\mbox{\scriptsize FERRO}}=0$ 
indicating a single solution. The  PARA free-energy  is obtained by 
plugging $\pi(x)=\delta(x)$ and $\widehat{\pi}(y)=\delta(y)$  into Eq.  (\ref{eq:freesym}):
\begin{eqnarray}
\label{eq:freepara} 
f_{\mbox{\scriptsize PARA}}&=& -\frac{1}{\beta}(\alpha \mbox{ ln}(\mbox{cosh }\beta) + \mbox{ln }2)
,\\
s_{\mbox{\scriptsize PARA}}&=&\alpha(\mbox{ln}(\mbox{cosh }\beta) -\beta\mbox{ tanh }\beta) +
\mbox{ln }2.
\end{eqnarray}
PARA solutions are unphysical  for
$\alpha > (\mbox {ln } 2)/(\beta \mbox { tanh }\beta - \mbox{ln ch }\beta)$, 
since the corresponding entropy is negative.
To complete the phase diagram picture we have to assess the spin-glass free-energy and
entropy. We  have seen in the beginning of this section  that  replica symmetric  SG and  PARA  solutions consist of the same field distributions for $K\rightarrow\infty$, 
implying unphysical behavior.  In order to produce a 
solution with  non-negative entropy one has to  break the replica symmetry. 
We use here  a  pragmatic way to build this solution, using the 
simplest one-step replica symmetry breaking known as {\it frozen spins}.
 
It was observed in Ref. \cite{gross} that for the REM  a  one-step 
symmetry breaking scheme gives the exact solution. In this scheme the $n$ 
replicas' space is divided to groups of $m$ identical solutions. 
It was shown that an abrupt transition in the order parameter from a unique 
solution (Edwards-Anderson parameter $q=1$, SG phase)  to a completely uncorrelated set of solutions ($q=0$, PARA phase) occurs. 
This transition takes place at a critical temperature
$\beta_g$ that can be found by solving the appropriate saddle-point equations;
 this temperature is given by the root of the replica symmetric entropy ($s_{\scriptsize RS}=0$) meaning that the RS-RSB transition occurs at the same point as the PARA-SG in this model. The  symmetry breaking parameter was found to be 
$m_g=\beta_g/\beta$, indicating  that this kind of solution 
is physical only for $\beta>\beta_g$, since 
$m_g \leq 1$ \cite{parisi}, 
indicating  a  PARA-SG phase transition. The free-energy can be computed  by plugging the order parameters in the effective
Hamiltonian, obtained  after averaging over the disorder  and taking the proper limits. It shows no dependence on the
temperature,  since for $\beta>\beta_g$ the system is completely
frozen in a single configuration.

For the Sourlas' code, in the regime we are interested in, SG solutions
to the saddle-point equations are given by $\pi(x)=\delta(x)$ and
$\widehat{\pi}(y)=\delta(y)$. The RSB-SG  free-energy that guaranties
continuity in the SG-PARA transition is identical to $f_{\mbox{\scriptsize PARA}}$,
since the SG and PARA solutions have exactly the same structure, to say:  

\begin{equation}
\label{eq:free_rsb_sg}
f_{\mbox{\scriptsize RSB-SG}}=-\frac{1}{\beta_g}\;(\alpha \;\mbox{ln}\,(\mbox{cosh }\,\beta_g) + 
\mbox{ln }\,2),
\end{equation}
where $\beta_g$ is a solution for $s_{\mbox{\scriptsize RS-SG}}= \alpha\;(\mbox{ln}\,(\mbox{cosh }\,\beta) -\beta\;\mbox{tanh }\,\beta) +\mbox{ln }\,2=0$. 

 In  Fig.\ref{phase}  we show  the phase diagram for a given code rate $R$ in
the temperature $T$ versus noise  level $p$  plane.

%@@@@@@@@@@@@@@@@@@@@@@@@@@@@@@@@@@@@@

\subsection{Shannon's Limit}
\label{sec:bound}
Shannon's analysis shows that up to a critical code rate $R_c$, which equals 
the channel capacity, it is possible to recover information with arbitrarily 
small  error probability for a given noise level. For the BSC :
\begin{equation}
\label{eq:shannon}
R_c=\frac{1}{\alpha_c}=1+p\mbox{ log}_2 \;p + (1-p)\mbox{ log}_2\; (1-p).
\end{equation}
 
Sourlas' code, in the case where  
$K\rightarrow\infty$ and  $C \sim {\cal O}(N^K)$ can be mapped onto  the  REM 
and  has been shown to be capable of  saturating Shannon's bound in the limit 
$R\rightarrow 0$ \cite{sourlas89}. In this section we extend the analysis
 to show that Shannon's bound can be attained by  Sourlas' 
code at zero temperature also for  
$K\rightarrow\infty$ limit but  with connectivity $C=\alpha K$.
In this limit the model is analogous to the diluted REM analyzed by Saakian
 in \cite {saakian}. 
The errorless phase is manifested in a FERRO phase with perfect 
alignment ($m=1$) (condition that is only possible for infinite $K$) up 
to a certain critical noise level; a further  noise level increase  produces 
frustration leading to a SG phase where
the misalignment is maximal ($m=0$). The FERRO-SG transition  is 
analogous to the transition from  errorless decoding to decoding with 
errors  described by  Shannon. A PARA phase is
also  present when the transmitted information  is insufficient to
recover the original  message ($R>1$).

% FIGURE 5
% Fields
\begin{figure}
\hspace*{2cm}
\epsfxsize=120mm  \epsfbox{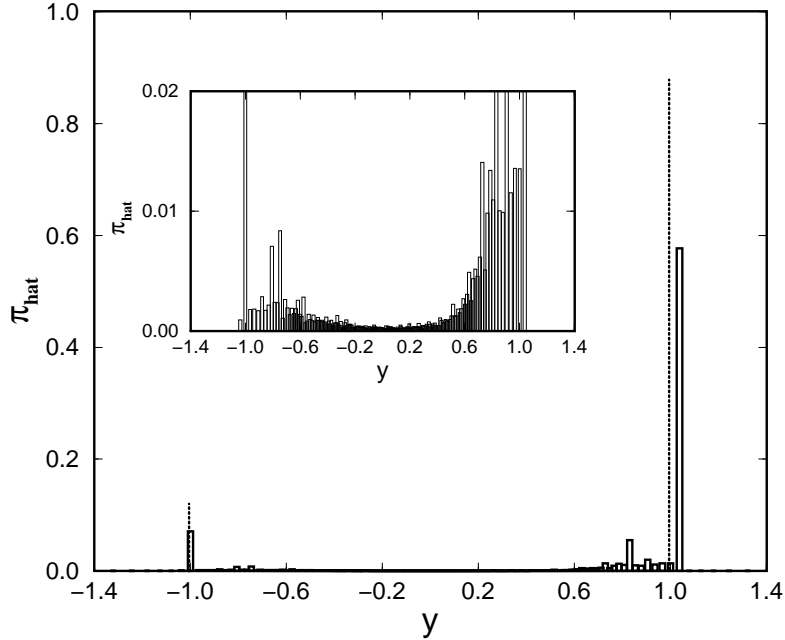}
\caption{Histogram representing the mean-field distribution
  $\widehat{\pi}(y)$ obtained by Monte-carlo integration at low 
temperature ($\beta=10$, $K=3$,$C=6$ and $p=0.1$). Dotted lines represent
solutions obtained by iterating self-consistent equations both with five peak and three peak ans\"atze. Inset: detailed view of the weak regular part arising in the Monte-carlo integration. }
\label{fields}
\end{figure}

At zero temperature   saddle-point equations (\ref{eq:saddle_point}) can
be rewritten  as:

\begin{eqnarray}
\label{eq:sp_infty}
\fl \pi(x) &=& \int \left[ \prod_{l=1}^{C-1} dy_{l} \ \widehat{\pi}(y_{l})\right]
 \ \delta \left[ x - \sum_{j=1}^{C-1} y_{j} \right] \\
\fl \widehat{\pi}(y) &=& \int \left[
\prod_{l=1}^{K-1} dx_{l} \ \pi(x_{l}) \right] \ \left\langle \delta 
\left[ y - \mbox{sign}(J \prod_{l=1}^{K-1} x_{l}) \mbox{min}(\mid J\mid,
 ... , \mid x_{K-1} \mid)\right] \right\rangle_{J} \ ,  
\nonumber
\end{eqnarray}

The solutions for these saddle-point equations may, in general, result
in probability distributions with singular and regular parts. As a
first approximation we choose the simplest  self-consistent  family
of solutions which are, since  $J=\pm 1$, given by:
\begin{eqnarray}
\widehat{\pi}(y)&=&p_+\delta(y-1)+p_0\delta(y)+p_-\delta(y+1)\\
\pi(x)&=&\sum_{l=1-C}^{C-1} T_{[p_{\pm},p_0;C-1]}(l) \,\delta(x-l), 
\end{eqnarray}
with
\begin{equation}
T_{ \left[ p_{+}, p_{0}, p_{-}; C-1 \right]} (l) = \sum^{\prime}_{ \{k,h,m\}} \frac{(C-1)!}{k! \ h! \ m!} \ p_{+}^{k} \
p_{0}^{h} \ p_{-}^{m},
\end{equation}
where the prime indicates that $k,h,m$ are such that $k-h=l; \ k+h+m=C-1$.
Evidence for this simple ansatz comes from  Monte-carlo integration of 
Equation (\ref{eq:saddle_point}) at  very low temperatures, that  shows  solutions comprising  three dominant peaks and  a relatively  weak regular part. 
Inside FERRO and PARA phases a more complex singular solution comprising 
five peaks $\widehat{\pi}(y)=p_{+2}\delta(y-1)+p_{+}\delta(y-0.5)+p_0\delta(y)+p_-\delta(y+0.5)+p_{-2}\delta(y+1)$ collapses back to the simpler three peak solution. In Fig.\ref{fields} we show a typical result of a  Monte-carlo integration for the field $\widehat{\pi}(y)$. The two peak that emerge by using either the three peak ansatz or the five peak ansatz are shown as dotted lines. In the inset we show the weak regular part of  the Monte-carlo solution.

 Plugging 
the above ansatz in the saddle-point equations one  can  write a  closed set
of equations in $p_{\pm}$ and $p_{0}$ that can be solved  numerically 
(see appendix D for details).

% FIGURE 6
% Zero temperature phase diagram 
\begin{figure}
\hspace*{2cm}
\epsfxsize=120mm  \epsfbox{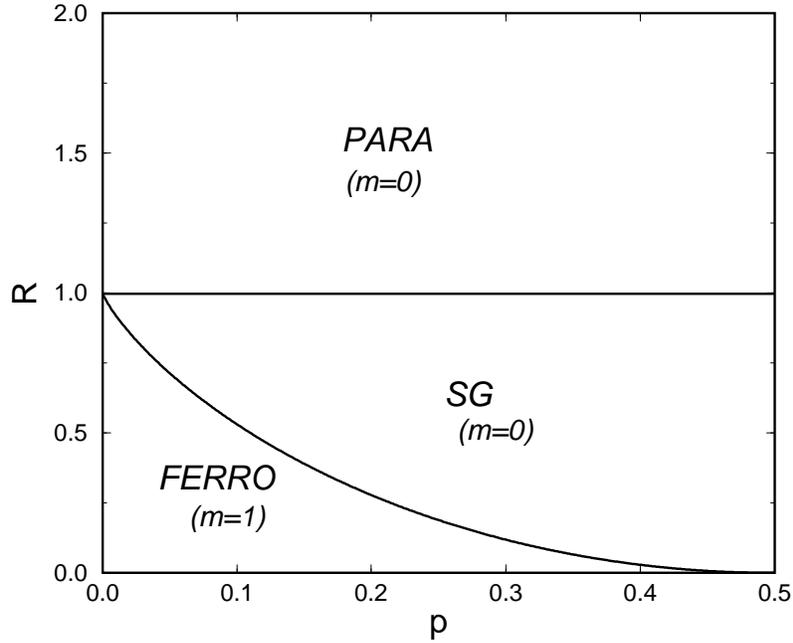}
\caption{Phase diagram in the plane code rate $R$ versus noise level $p$ 
for $K\rightarrow\infty$ and $C=\alpha K$ at zero 
temperature. The FERRO-SG coexistence line  corresponds to the Shannon's 
bound.}
\label{bound}
\end{figure}
 
The three peak solution can be of three types: FERRO 
($p_+>p_-$), PARA ($p_0=1$) or  SG ($p_-=p_+$). 
Computing  free-energies and entropies enables one to construct the phase
 diagram. At zero temperature the PARA free-energy is $f_{\mbox{\scriptsize PARA}}=-\alpha$ 
and the entropy is $s_{\mbox{\scriptsize PARA}}=(1-\alpha)\mbox{ ln }2$, this  phase is physical
only for $\alpha<1$, what is expected  since it corresponds exactly to the 
regime where  the transmitted information is not sufficient to recover 
the actual message ($R>1$).

The FERRO free-energy does not depend on the temperature, having
the form $f_{\mbox{\scriptsize FERRO}}=-\alpha(1-2p)$ with  entropy  $s_{\mbox {\scriptsize FERRO}}=0$. One can 
find the 
FERRO-SG coexistence line  that corresponds to the maximum performance of a 
Sourlas' code by equating 
Eq.(\ref{eq:free_rsb_sg}) and $f_{\mbox{\scriptsize FERRO}}$. 
Observing that $\beta_g=\beta_N(p_c)$
(as seen in  Fig.\ref{phase} ) we found that this transition 
coincides with  Shannon's bound 
Eq.(\ref{eq:shannon}). It is interesting to note that in the large $K$ regime
both  RS-FERRO  and RSB-SG free-energies (for $T<T_g$)
do not depend on the temperature, it means that Shannon's bound
is valid also for finite temperatures up to $T_g$. 
In  Fig.\ref{bound} we give the complete zero temperature phase diagram. 

The stability of  replica 
symmetric FERRO and PARA solutions used to obtain Shannon's bound can
be checked  using Eq.(\ref{eq:stability}) at zero temperature:

\begin{equation}
\label{eq:stability0temp}
\fl \lim_{\beta \rightarrow \infty}
\frac{1}{\beta} \int \left[\prod_{l=1}^{K-2} dx_{l} \ 
\pi(x_{l}) \right] \left\langle
 \ln \left[ 1+ \tanh \beta J \ \tanh^2 \beta x\prod_{j=1}^{K-2} \tanh
 \beta x_{j} \right]
\right\rangle_{J} \ge 0, 
\end{equation}
for all $x$.
For PARA solutions the above integral vanishes, trivially satisfying the
condition, while for FERRO solution in the $K$ large regime, 
$x_l\approx{\cal O}(K)$ and the integral becomes
\begin{equation}
-2p \left [ \left (1- \Theta \left( x+1 \right ) \right )
+ |x| \left (\Theta \left( x+1 \right ) -\Theta \left( x- 1 \right ) \right )
+ \Theta \left( x- 1 \right ) \right ], 
\end{equation}
where $\Theta(x)=1$ for $x\geq 0$ and $0$ otherwise,
indicating instability for $p>0$. 
For the noiseless
case $p=0$ the stability condition is satisfied. The instability of  
FERRO phase opens the possibility that Sourlas' code does not saturate 
Shannon's bound, since a correction to  the
FERRO solution could change  FERRO-SG transition line. However, it was 
shown in Section \ref{sec:symmetric} that this instability vanishes for large
temperatures, what supports, to some extent,  the FERRO-SG line obtained and the saturation 
of  Shannon's bound in some region, as long  as the temperature is  lower than  Nishimori's temperature.
For finite temperatures the stability condition for FERRO solution
can be rewritten as:
\begin{equation}
\left(1+\mbox{tanh}(\beta)\mbox{tanh}^2(\beta x)\right)^{(1-p)}
\left(1-\mbox{tanh}(\beta)\mbox{tanh}^2(\beta x)\right)^p \ge 1 \; \forall x.
\end{equation}
For $p=0$ the condition is clearly  satisfied. For finite $p$ 
a critical temperature above which the stability condition is fulfilled can be found numerically.
In Fig.\ref{phase} we show this critical temperature in the phase diagram;
 one can see that there is a considerable region in which our result that 
 Sourlas' code can saturate  Shannon's bound is supported. Conclusive evidence to 
that will be given by  simulations presented in Section \ref{sec:decoding}.

%@@@@@@@@@@@@@@@@@@@@@@@@@@@@@

\subsection{Finite K Case}
 \label{sec:finite}
 Although  Shannon's bound only can be attained in the limit 
 $K\rightarrow\infty$, it was shown in the Section \ref{sec:Kinfty} 
 that  there are some possible drawbacks, mainly  in the decoding of 
 messages  encoded by large $K$ codes, due to  large 
barriers which are expected to occur between  PARA   and  FERRO
states.   In this section we consider the finite $K$  case, 
for which  we can solve the  RS saddle-point equations (\ref{eq:saddle_point}) 
for arbitrary 
temperatures using Monte-carlo integration. We can also obtain solutions
for the zero temperature case using the simple iterative method described in
Section \ref{sec:bound}. 
 
% FIGURE 7
% K finite
\begin{figure}
\hspace*{2cm}
\epsfxsize=120mm  \epsfbox{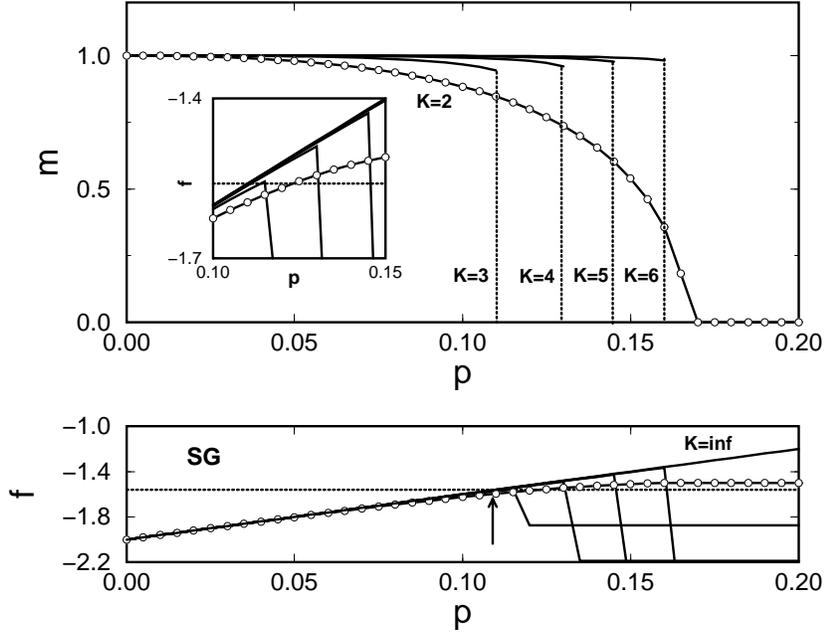}
%\vspace{1cm}
\caption{Top: zero temperature magnetization $m$ as a function of the noise level $p$ for various $K$ values at code rate $R=1/2$, as
 obtained by the iterative method  
. Notice that  the  RS theory predicts a transition
 of second order for  $K=2$  and  first  order for $K>2$. Bottom:
RS-FERRO free-energies (white circles for $K=2$ and from the 
left: $K\,=\,3,4,5$ and $6$) and RSB-SG free-energy (dotted line) as 
functions of the noise level $p$. 
The arrow indicates the region where the RSB-SG phase starts to dominate.
 Inset: a detailed view of the RS-RSB transition region.}
\label{kfinite}
\end{figure}

At the top of Fig.\ref{kfinite} we show  the zero temperature 
magnetization $m$ as a function of  the noise level  $p$ at 
code rate $R=1/2$.
 These curves were obtained by using the  
three peak ansatz of the Section \ref{sec:bound}. 
It can be seen that the transition is of second order for $K=2$ and first 
order for $K>3$ similarly to extensively connected models. The transition
as described by the RS solution tends to $p=0.5$ as $K$ grows. Note that
this does not correspond to perfect retrieval since the RSB spin glass
phase dominates for $p>p_c$ (see bottom of Fig.\ref{kfinite}). In the bottom 
figure we plot  RS free-energies and  RSB frozen 
spins free-energy, from which we determine the 
critical probability $p_c$ where the transition occurs (pointed by an arrow).
 After the transition, free-energies for $K=3,4,5$ and $6$ acquire values
that are lower than the SG free-energy; nevertheless, the entropy 
is negative and these free-energies are therefore unphysical. It is 
remarkable that this critical value does  not change significantly
for finite $K$  in comparison to  infinite $K$.
Observe that  Shannon's bound
cannot be attained for finite $K$, since $m=1$ exactly only if 
$K\rightarrow\infty$.

The  $K=2$ model with extensive connectivity (SK)  is known to be 
somewhat special, a full Parisi solution
is needed to recover the concavity of the  free-energy and
the Parisi order function has a continuous behavior \cite{mezard}.
No stable solution
is known for the intensively connected model (Viana-Bray model). 
In order to check the theoretical result obtained one relies  on 
simulations of the decoding process  
at low temperatures. In Section VIII we show that the simulations are in good 
agreement with the theoretical results.

%@@@@@@@@@@@@@@@@@@@@@@@@@@@@@@

\subsection {Gaussian Noise}
\label{sec:gauss}
Using the replica symmetric free-energy  (\ref{eq:freesym}) and the 
frozen spins
RSB  free-energy (\ref{eq:free_rsb_sg})  one can easily extend the analysis
to other noise types. The general PARA free-energy and entropy can be written:
\begin{eqnarray}
\label{eq:fparagen}
f_{\mbox{\scriptsize PARA}}&=&-\frac{1}{\beta}\;\left(\alpha\;\langle \mbox{ln}\,(\mbox{ch }\beta J)\rangle_J +\mbox {ln }\,2\right) \nonumber \\
s_{\mbox{\scriptsize PARA}}&=&\alpha\;\left(\langle \mbox{ln}\,(\mbox{ch }\beta J)\rangle_J - \beta \langle J\;\mbox{tanh}\,(\beta J)\rangle_J\right)+\mbox {ln }2. 
\end{eqnarray} 
The SG-RSB free-energy is given by :
\begin{equation}
\label{eq:fsggen}
f_{\mbox{\scriptsize SG-RSB}}=-\frac{1}{\beta_g}\;\left(\alpha\;\langle \mbox{ln }(\mbox{ch }\beta_g J)\rangle_J +\mbox {ln }2 \right), 
\end{equation} 
with $\beta_g$ defined as  the solution of
\begin{equation}
\label{eq:entropy}
\alpha\;\left(\langle \mbox{ln }(\mbox{ch }\beta_g J)\rangle_J - \beta_g 
\langle J\;\mbox{tanh }(\beta_g J)\rangle_J\right)+\mbox {ln }2 =0 .
\end{equation}

The FERRO free-energy is in general given by  
$f_{\mbox{\scriptsize FERRO}}=-\alpha\;\langle J\rangle_J=-\alpha\;\langle J \; 
\mbox{tanh }(\beta_N J)\rangle_J$ (see Appendix \ref{app:nishifree}).
The maximum performance of the code is  defined by the critical line :
\begin{equation}
\label{eq:line}
\alpha\left(\langle \mbox{ln}(\mbox{ch }\beta_g J)\rangle_J - \beta_g 
\langle J\;\mbox{tanh}(\beta_N J)\rangle_J\right)+\mbox {ln }2 =0,
\end{equation}
obtained by equating  free-energies in PARA and FERRO phases. 
Comparing this expression with  entropy (\ref{eq:entropy}) it can be
seen that $\beta_g=\beta_N$ at the critical line; the 
same behavior observed in the BSC case. From Eq.(\ref{eq:line}) one can 
write:
\begin{equation}
\label{eq:capacity}
R_c=\beta_N^2 \frac{\partial}{\partial \beta}
\left[\frac{1}{\beta}\langle \mbox{log}_2 
\mbox{ cosh}(\beta J)\rangle_J\right]_{\beta=\beta_N},
\end{equation}
that can be used to compute the performance of the code for arbitrary 
symmetric noise. 

%FIGURE 8
% Gaussian noise
\begin{figure}
\hspace*{2cm}
\epsfxsize=120mm  \epsfbox{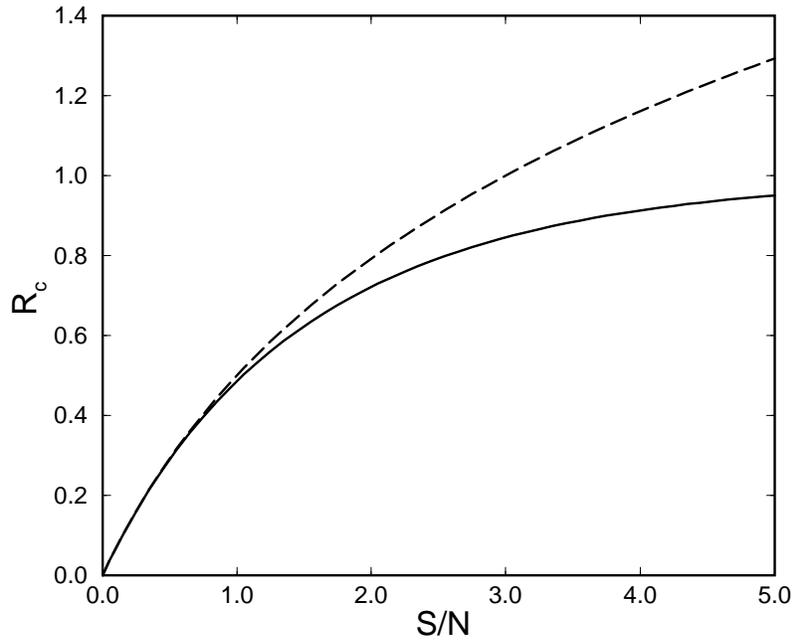}
\vspace{1cm}
\caption{Critical code rate $R_c$ and channel capacity  for a binary Gaussian 
channel  as a function of the signal to noise rate $S/N$ (solid line). 
Sourlas' code saturates  Shannon's bound. Channel capacity  of the 
unconstrained  Gaussian channel (dashed line).}
\label{gaussian}
\end{figure}

Supposing that the encoded bits can acquire totally unconstrained values 
Shannon's bound for Gaussian noise is given by 
$R_c=\frac{1}{2}\mbox{ log}_2(1 +S/N)$, where $S/N$ is the signal to 
noise ratio, defined as the ratio of source energy per bit 
(squared amplitude) over the spectral density of the noise (variance). 
If one constrains the encoded bits to binary values $\{\pm 1\}$ the capacity of a Gaussian  channel
is:
\begin{equation}
\label{eq:capacity_gauss}
R_c=\int dJ\;P(J\mid 1) \mbox{ log}_2 P(J\mid 1) - \int dJ \; P(J)
\mbox{ log}_2 P(J),  
\end{equation}
where $P(J\mid J^0)=\frac{1}{\sqrt{2\pi \sigma^2}}\mbox{ exp}(-\frac{(J-J^0)^2}{2\sigma^2})$.

In Fig.\ref{gaussian} we show  the performance of Sourlas' code
in  a Gaussian  channel together with the capacities of the 
unconstrained and binary Gaussian channels. We show that 
 $K\rightarrow\infty$, $C=\alpha K$ Sourlas' code
saturates  Shannon's bound for the binary Gaussian channel as well.
The significantly lower performance in  the  unconstrained 
Gaussian channel can be trivially explained by the binary coding
scheme while  signal and noise are allowed to acquire real values.
%@@@@@@@@@@@@@@@@@@@@@@@@@@@@@@@@@@@@@@@@@@@@@@@@@@@@@@@@@@@@@@@@@@@

\section {Decoding Dynamics}
\label{sec:decoding}

%@@@@@@@@@@@@@@@@@@@@@@@@@@@@@@
\subsection{Belief Propagation}
\label{sec:belief}

\begin{figure}
\hspace*{4cm}
\epsfxsize=80mm  \epsfbox{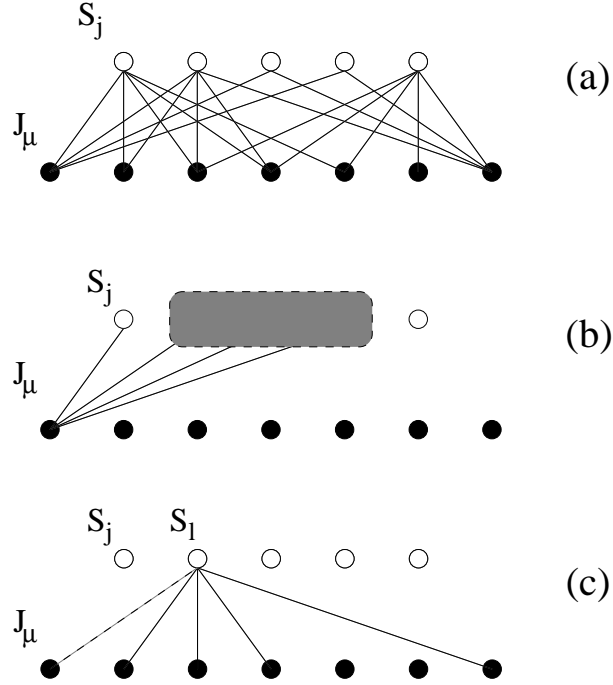}
\vspace{1cm}
\caption{(a) Belief network representing an error-correcting code. 
Each bit $S_j$ (white circles) is linked to exactly $C$ checks and 
each check (black circles) $J_{\mu}$ is linked to exactly $K$ bits. 
(b) Graphical representation of the field $r_{\mu j}$. The grey box
represents the mean field contribution $\prod_{l\in {\cal L}(\mu)\setminus j}q_{\mu l}$ 
of the other bits on the check $J_\mu$. (c) Representation of one of the
fields $q_{\mu l}$. }
\label{belief}
\end{figure}

  The decoding process of an error-correcting code relies on
computing averages over the marginal posterior probability 
$P(S_j\mid \mbox{\boldmath $J$})$ for each one 
of the $N$ message bits $S_j$  given the corrupted encoded bits  
$J_\mu$ (checks), where $\mu=\langle i_1 \ldots i_K \rangle$ is one of 
the $M$ sets chosen by the tensor ${\cal A}_\mu$. 
The  probabilistic dependencies 
existing  in the code can be represented as a bipartite graph known as a 
{\it belief  network} where  nodes in one layer correspond to the $M$ 
checks $J_\mu$ while  nodes in the  other to the $N$ bits $S_j$.
 Each check is connected 
to exactly $K$ bits and each bit is connected exactly to $C$ checks 
(see Fig.\ref{belief}a).

Pearl \cite{pearl} proposed an iterative algorithm for computation of 
marginal probabilities in belief networks. These algorithms operate by 
updating beliefs (conditional probabilities) locally and  propagating them. 
Generally the  convergence of these iterations  
 depends on the absence of loops in the graph. As can be seen
in Fig.\ref{belief}a, networks that define  error-correcting
codes may include  loops and convergence problems may occur. 
Recently it was shown that in some cases  Pearl's algorithm works
even in the presence of loops \cite{weiss}.

% FIGURE 9
% Belief

The particular use  of  belief networks as decoding 
algorithms   for error-correcting codes based
on  sparse matrices was discussed  by MacKay in \cite{mackay95b}. 
In this work  a  loop-free approximation  for the graph in Fig.\ref{belief}a 
was proposed (see  \cite{pearl} for a general discussion
on such approximations). In fact, it was shown in \cite{urbanke} that the probability of finite length loops in these graphs vanishes with the system size. 

In this framework the
network is decomposed in a way to avoid loops and the conditional probabilities
  $q^{(S)}_{\mu j}$ and $r^{(S)}_{\mu j}$ are  computed.
 The set of bits in a check $\mu$ is defined as ${\cal L}(\mu)$ and the set of checks over the bit $j$ as ${\cal M}(j)$. The probability that  $S_j=S$
 given information on all checks other than $\mu$ is denoted
 $q^{(S)}_{\mu j} =P(S_j=S\mid \{J_{\nu}:\nu \in {\cal M}(j)\setminus\mu\})$ 
 and $r^{(S)}_{\mu j} = \mbox{Tr}_{\{S_l:l\in{\cal L}(\mu)\setminus j\}}
P(J_{\mu}\mid S_j=S, \{S_l:l \in {\cal L}(\mu)\setminus j \}) 
\prod _{l\in {\cal L}(\mu)\setminus l} q^{(S_l)}_{\mu l}$ is the probability of the check $J_{\mu}$ if the bit $j$ is fixed to $S_j=S$ and the other bits 
involved are supposed to have distributions given by $q^{(S_i)}_{\mu i}$
. In  Fig.\ref{belief}b one can see a  graphical representation of
$r^{(S)}_{\mu j}$ that can be interpreted as the influence of the bit $S_j$ and the  mean-field $\prod _{l\in {\cal L}(\mu)\setminus l} q^{(S_l)}_{\mu l}$ 
(representing bits in  ${\cal L}(\mu)$ over than $l$)  over the check $J_\mu$. In the Fig.\ref{belief}c 
we see that each field $q^{(S)}_{\mu l}$ represents  the influence of the 
checks in  ${\cal M}(l)$,  excluding $\mu$, over each bit $S_l$, this 
setup excludes  the loops that may exist in the actual network.

Employing Bayes theorem, $q^{(S)}_{\mu j}$ can be rewritten as:
\begin{equation}
q^{(S)}_{\mu j} = a_{\mu j}\;P(\{J_{\nu}:\nu \in {\cal M}(j)\setminus\mu\}\mid S_j)\;p^{(S)}_{j},
\end{equation}
where $a_{\mu j}$ is a normalization constant such that 
$q^{(+1)}_{\mu j}+q^{(-1)}_{\mu j}=1$ and $p^{(S)}_{j}$ is the prior
probability over the bit $j$. The distribution $P(\{J_{\nu}:\nu \in {\cal M}(j)\setminus\mu\}\mid S_j)$ can be replaced by a  mean-field approximation by 
factorizing dependencies using  fields  $r^{(S)}_{\mu j}$:
\begin{eqnarray}
\label{eq:belief}
q^{(S)}_{\mu j}&=&a_{\mu j}p^{(S)}_{j}\prod_{\nu\in{\cal M}(j)\setminus\mu}
r^{(S)}_{\nu j} \nonumber \\
r^{(S)}_{\mu j} &=& \mbox{Tr}_{\{S_l:l\in{\cal L}(\mu)\setminus j\}}
P(J_{\mu}\mid S_j=S, \{S_i:i \in {\cal L}(\mu)\setminus j \}) 
\prod _{i\in {\cal L}(\mu)\setminus j} q^{(S_i)}_{\mu i}.
\end{eqnarray}

A message estimate
$\widehat\xi_j =\mbox{sign}\left(\langle S_j \rangle_{q^{(S)}_j}\right)$
can be obtained by  solving the above equations and
computing the pseudo-posterior:
\begin{equation}
q^{(S)}_{j}=a_{j}p^{(S)}_j\prod_{\nu\in{\cal M}(j)}r^{(S)}_{\nu j},
\label{eq:pseudo}
\end{equation}
where $a_{j}$ is a normalization constant. 
 
By  taking advantage of the normalization conditions for the distributions
$q^{(+1)}_{\mu j}+q^{(-1)}_{\mu j}=1$ and  
$r^{(+1)}_{\mu j}+r^{(-1)}_{\mu j}=1$
one can change variables and reduce the number of equations (\ref{eq:belief}) to  the couple
$\delta q_{\mu j} = q^{(+1)}_{\mu j}-q^{(-1)}_{\mu j}$ and 
$\delta r_{\mu j} = r^{(+1)}_{\mu j}-r^{(-1)}_{\mu j}$.
Solving these equations, one  can find back $r^{(S)}_{\mu j}=\frac{1}{2}
\left(1\;+\;\delta r_{\mu j}S_j \right)$ and the
pseudo-posterior can be calculated to obtain the estimate.

%@@@@@@@@@@@@@@@@@@@@@@@@@@@@@@@@@@@@
\subsection{Connection with Statistical Physics}
\label{sec:connection}

The belief propagation algorithm was shown in  \cite{mackay95b} to outperform 
other methods  such as simulated annealing. In \cite{ks98a} it was proposed
that this framework can be reinterpreted  using statistical physics. 
The main ideas behind the   approximations  contained in (\ref {eq:belief})  are somewhat similar to the  Bethe \cite{bethe} approximation to diluted 
two-body spin glasses. Actually, 
for systems involving two-body interactions it is  known that the Bethe
approximation is equivalent to solving exactly a model defined on a 
Cayley tree and that this  is a good approximation
for finitely connected systems in the thermodynamical limit \cite{wong_d}.
In fact, loops in the connections  become rare as the system size grows 
and can be neglected without introducing significant errors. The belief propagation can be seen as a Bethe-like approximation for multiple bodies interaction systems.

The mean-field approximations used here are also quite similar 
to the TAP approach  \cite{tap}. The fields $q^{(S)}_{\mu j}$ correspond
to the mean influence of other sites other the site $j$ and  
the fields  $r^{(S)}_{\nu j}$ represent the influence of $j$ back over the system (reaction fields). 

The analogy can be exposed by  observing  that the likelihood 
$p(J_\mu\mid \mbox{\boldmath $S$})$ 
is proportional to the Boltzmann weight:
\begin{equation}
w_B(J_\mu \mid \{S_j:j \in {\cal L}(\mu) \}) = \mbox{exp}\left(-\beta  J_\mu  \; \prod_{i\in\mu} S_{i}\right).
\end{equation} 
That can be also written in the more convenient form:
\begin{equation}
\label{eq:likelihood}
w_B(J_\mu \mid \{S_j :j \in {\cal L}(\mu) \}) = \frac{1}{2} \mbox{cosh}(\beta J_\mu)\left( 1\; + 
\;\mbox{tanh}(\beta J_{\mu})\; \prod_{j\in{\cal L}(\mu)}S_j \right).
\end{equation} 

The variable $r_{\mu j}^{(S_j)}$  can then be seen as proportional to the effective   Boltzmann weight  obtained by fixing the bit $S_j$:
\begin{equation}
\label{eq:effective}
w_{\mbox{\scriptsize eff}}(J_\mu \mid S_j) 
= \mbox{Tr}_{\{S_l\; :\;l \in {\cal L}(\mu)\setminus j \}}\; 
w_B(J_\mu \mid \{S_l\; :\;l \in {\cal L}(\mu) \})\prod _{l\in {\cal L}(\mu)\setminus j} 
 q^{(S_l)}_{\mu l}. 
\end{equation} 
 Plugging  Eq.(\ref{eq:likelihood}) for the likelihood in equations
(\ref{eq:belief}), using the fact that the prior probability is given by $p_j^{(S)}=\frac{1}{2}\left(1+\mbox{tanh}(\beta S F)\right)$ and computing  $\delta q_{\mu j}$ and $\delta r_{\mu j}$:

\begin{eqnarray}
\label{eq:tap}
\delta r_{\mu j}&=&\mbox{tanh}(\beta J_\mu) \prod_{l\in{\cal L}(\mu)\setminus j} \delta q_{\mu l}\nonumber \\
\delta q_{\mu j}&=&\mbox{tanh}\left(\sum_{\nu\in{\cal M}(l)\setminus \mu} 
\mbox{tanh}^{-1}( \delta r_{\nu j}) +\beta F \right).
\end{eqnarray}

% FIGURE 10
% TAP K=2
\begin{figure}
\hspace*{2cm}
\epsfxsize=120mm  \epsfbox{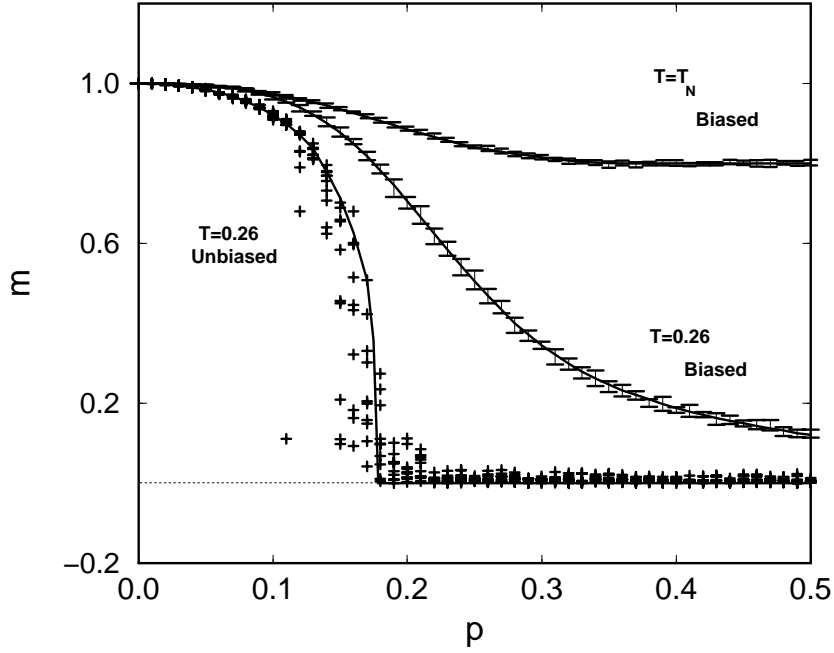}
%\vspace{1cm}
\caption{Magnetization as a function of the flip probability $p$ for 
decoding using TAP equations for $K=2$. From the bottom: 
Monte-carlo solution of the RS saddle-point equations 
 for unbiased messages ($f_s=0.5$) at $T=0.26$  (line) and 
$10$ independent runs of TAP decoding for each flip probability (plus signs),
 $T=0.26$ and biased messages ($f_s=0.1$) at 
 Nishimori's temperature $T_N$.}
\label{TAP1}
\end{figure}

The pseudo-posterior can then be calculated:
\begin{equation}
\label{eq:pseudoposterior}
\delta q_{j}=\mbox{tanh}\left(\sum_{\nu\in{\cal M}(l)}
\mbox{tanh}^{-1}( \delta r_{\nu j}) +\beta F \right),
\end{equation}
providing  Bayes' optimal decoding 
$\widehat{\xi}_j=\mbox{sign}(\delta q_{j})$. It is 
important at this point  to support the mean-field 
assumptions used here  by methods  of  statistical 
physics \cite{ks98a}. The factorizability of the probability 
distributions can be explained  by weak correlations between connections 
(checks) and by the cluster property: 
\begin{equation}
\lim_{N\rightarrow \infty} \frac{1}{N^2}\sum_{i\neq j}
\left(\langle S_i  S_j\rangle_{p(S\mid J)} -
\langle S_i \rangle_{p(S\mid J)}
\langle S_j \rangle_{p(S\mid J)}\right)^2 \rightarrow 0
\end{equation}
that  bits $S_j$ obey within a pure state \cite{mezard}.

% FIGURE 11
% TAP K=5
\begin{figure}
\hspace*{2cm}
\epsfxsize=120mm  \epsfbox{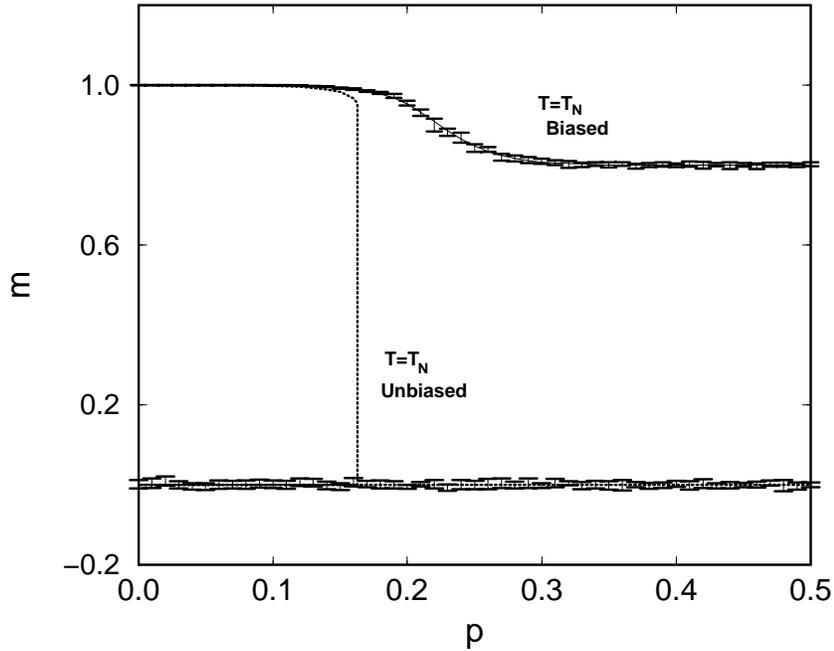}
%\vspace{1cm}
\caption{Magnetization as a function of the flip probability $p$ for 
decoding using TAP equations for $K=5$. 
The dotted line is the replica symmetric saddle-point equations Monte-carlo
integration for unbiased messages ($f_s=0.5$) at the Nishimori's temperature $T_N$. The 
bottom error bars correspond to $10$ simulations using the TAP decoding. 
The decoding performs badly on average in this scenario. The upper curves
are for biased messages ($f_s=0.1$) at the Nishimori's temperature $T_N$. 
The simulations agree with results obtained using the replica symmetric ansatz
and  Monte-carlo integration.}
\label{TAP2}
\end{figure}

One can push the above connections even further. Eqs.(\ref{eq:tap}), of course, depend on the particular received message $\mbox{\boldmath $J$}$. In order to  make the analysis  message independent, one can use 
a gauge transformation $\delta r_{\mu j} \mapsto \xi_j \delta r_{\mu j}$ and
$\delta q_{\mu j} \mapsto \xi_j \delta q_{\mu j}$  to write:
\begin{eqnarray}
\label{eq:gaugetap}
\delta r_{\mu j}&=&\mbox{tanh}(\beta J) \prod_{l\in{\cal L}(\mu)\setminus j} \delta q_{\mu l}\nonumber \\
\delta q_{\mu j}&=&\mbox{tanh}\left(\sum_{\nu\in{\cal M}(l)\setminus \mu} 
\mbox{tanh}^{-1}( \delta r_{\nu j}) +\beta \xi_j F \right).
\end{eqnarray}
In this form a success in the decoding process correspond to $\delta r_{\mu j}>0$ and $\delta q_{\mu j}=1$ for all $\mu$ and $j$. For a large number of iterations, one can expect the ensemble of  belief networks to converge to an equilibrium distribution where  $\delta r$ and $\delta q$ are random variables sampled from distributions $\widehat{\rho}(y)$ and $\rho(x)$ respectively.
By transforming these variables as   $\delta r=\mbox{tanh}(\beta y)$ and 
$\delta q=\mbox{tanh}(\beta x)$ and considering the actual message and noise as quenched disorder,  Eqs.(\ref{eq:gaugetap}) can be rewritten as:
\begin{eqnarray}
\label{eq:newgaugetap}
y&=&\frac{1}{\beta}\left\langle \mbox{tanh}^{-1}\left(\mbox{tanh}(\beta J) \prod_{j=1}^{K-1} \mbox{tanh}(\beta x_j)\right)\right\rangle_J\nonumber \\
x&=&\left\langle \sum_{j=1}^{C-1} y_j + \xi F\right\rangle_\xi.
\end{eqnarray}

The above relations lead to a dynamics on the distributions $\widehat{\rho}(y)$ and $\rho(x)$, that is exactly the same obtained when solving iteratively 
 RS saddle-point equations (\ref{eq:saddle_point}). The probability distributions  $\widehat{\rho}(y)$ and $\rho(x)$ can be ,therefore, identified with $\widehat{\pi}(y)$ and $\pi(x)$ respectively and the RS solutions correspond to decoding  a generic message using belief propagation averaged over an ensemble of different codes, noise and signals.

Eqs.(\ref{eq:tap}) are now used to show the agreement between the simulated 
decoding and  analytical calculations. 
For each run, a fixed
code is used to generate  20000 bit codewords from 10000 bit messages,
corrupted versions of the codewords are then decoded using  (\ref{eq:tap}).
Numerical solutions for 10 individual 
runs are presented in Figs.\ref{TAP1} and \ref{TAP2},
initial conditions are chosen as $\delta r_{\mu l}=0$
 and $\delta q_{\mu l}=\mbox{tanh}(\beta F)$ reflecting  prior 
 beliefs.  In  Fig.\ref{TAP1} 
we show results for $K=2$ and  $C=4$
in the unbiased case, at code rate $R=1/2$
(prior probability $p_j^{(1)}=f=0.5$)
 at a low temperature  $T=0.26$ (we avoided $T=0$ due to numerical 
difficulties). 
Solving  saddle-point equations (\ref{eq:saddle_point}) numerically using 
Monte-carlo integration methods  we obtain solutions with good agreement to 
simulated decoding. In the same figure we show the performance for the case of 
biased messages ($p_j^{(1)}=f_s=0.1$), at code rate $R=1/4$. Also here the 
agreement with Monte-carlo integrations is rather convincing. The third 
curve in  Fig.\ref{TAP1} shows the performance 
for  biased  messages at Nishimori's temperature $T_N$, as expected, it
is far superior compared to low temperature performance and the 
agreement with Monte-carlo results is even better.

In Fig.\ref{TAP2} we show the results obtained  for $K=5$ and $C=10$.
For unbiased messages the system is extremely
sensitive to the choice of  initial conditions and does not perform
well in average even at Nishimori's temperature. For biased messages
($f_s=0.1$, $R=1/4$) results are far better and in agreement
with Monte-carlo integration of the RS saddle-point equations.

The experiments show that  belief propagation methods may be used 
successfully for decoding Sourlas-type codes in practice, and provide 
solutions that are well described by  RS analytical solutions. 

%@@@@@@@@@@@@@@@@@@@@@@@@@@@@@@@@@@@

\subsection{Basin of Attraction}
\label{sec:basin}

% FIGURE 12
% Basins of attraction
\begin{figure}
\hspace*{2cm}
\epsfxsize=120mm  \epsfbox{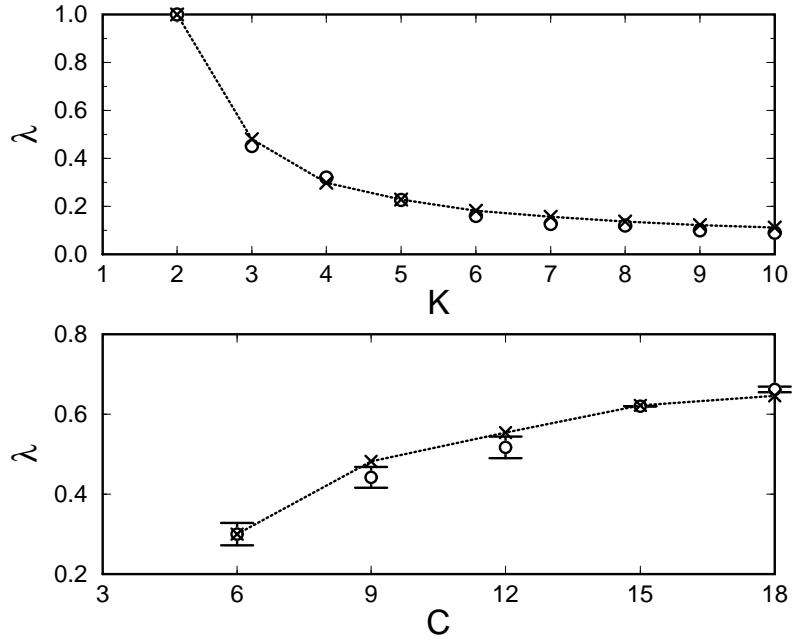}
%\vspace{1cm}
\caption{Top: Maximum initial deviation $\lambda$ for decoding. Top: $\lambda$
as function of  the number of interactions $K$. Circles are 
averages over $10$ different codes with $N=300$, $R=1/3$ and noise level 
$p=0.1$.  Bottom: $\lambda$ as function of the connectivity $C$. 
Circles are averages over $10$ codes with $N=300$, $K=3$ 
and noise level $p=0.1$. Lines and $\times$'s correspond to the RS dynamics 
described by the saddle-point equations. }
\label{basins}
\end{figure}

To asses the size of the  basin of attraction we consider the decoding process as a dynamics in the graphs space where  edges
$\delta q_{\mu j}$  are  considered as 
dynamical variables. In  gauged transformed equations (\ref{eq:gaugetap})
, the perfect decoding of a message correspond to $\delta q_{\mu j}=1$ .
 To analyse the basin of attraction we start with random initial values 
with a given normalized deviation from the perfect decoding  $\lambda=\frac{1}{NC}\sum_{\mu j}(1-\delta q_{\mu j}^{0})$. It is analogous to the finite magnetizations used in the naive mean-field of Section II, since a given $\delta q_{\mu j}^{0}$ corresponds to a given magnetization value by using Eq.(\ref{eq:pseudoposterior}).

In  Fig.\ref{basins} we show the maximal deviation in initial conditions required for successful decoding. 
Top figure shows an  average over $10$ different codes with
 $N=300$ (circles) for a fixed code rate $R=1/3$, fixed noise level $p=0.1$ and
increasing $K$. Bottom figure shows the maximal deviation in initial conditions  for a fixed number of spins per interaction $K=3$, noise level $p=0.1$ and increasing  $C$. We confirm the fidelity of the RS description by comparing the experimental results with the basin of attraction predicted by  saddle-point equations (\ref{eq:saddle_point}). One can interpret these
equations as a dynamics in the space of  distributions $\pi(x)$.
Performing the transformation $X=\mbox{tanh}(\beta x)$, one can move to
the space of distributions $\Pi(X)$ with support over $[-1,+1]$.
 The initial conditions  can then be  described simply as $\Pi^0(X)=(1-\frac{\lambda}{2})\delta(X-1)+\frac{\lambda}{2}\delta(X+1)$. In Fig.\ref{basins}
we show the basin of attraction of this dynamics as lines and $\times$'s. 

The $K=2$ case is the only practical code  from a dynamical point 
of view, since it has the largest basin of attraction and no  prior knowledge on the message is necessary for decoding. Nevertheless, this code's performance degrades  faster than the $K>2$ case as shown in Section III, which points to a compromise between good dynamical properties in one side and good performance in the other. One idea could be having a code with changing $K$, starting with $K=2$ to guarantee convergence and progressively increasing its values to improve 
the performance \cite{idosaad}.

On the other hand, the basin of attraction increases with  $C$. Again  it 
points to a trade off between good equilibrium properties 
(small $C$ and large code rates) and good dynamical properties (large $C$, 
large basin of attraction). Mixing small and large $C$ values in the same code 
seems to be a way to take advantage of this trade-off \cite{LMSS,davey,VSK}.

%@@@@@@@@@@@@@@@@@@@@@@@@@@@@@@@@@@@@@@@@@@@@@@@@@@@@@@@@@@@@@@@@@@@@

\section{Concluding Remarks}

In this paper we studied, using the replica
approach, a finite connectivity many-body spin glass that
corresponds to  Sourlas' codes for finite code rates. We have shown, 
using a simplified one step RSB solution for 
spin glass phase, that for  $K\rightarrow\infty$ and  $C=\alpha K$ 
regime at low temperatures the system exhibits a FERRO-SG phase transition 
that corresponds to Shannon's bound. However, we have also shown that the
decoding problem for large $K$ has bad convergence properties
when simulated annealing strategies are used.    

We  were able to find replica symmetric solutions for finite $K$ and 
found good agreement with practical decoding performance using belief
networks. Moreover, we have shown that  RS saddle-point equations actually 
describe the mean behavior of  belief propagation algorithms.

We studied the dynamical properties of belief propagation and compared to 
 statistical physics predictions, confirming the validity of the description. The basin of attraction was shown to depend on  $K$ and $C$. Strategies for improving the performance were discussed.

 The same methodology has been recently employed successfully \cite{kms99}
to state-of-the-art algorithms as the recent rediscovered Gallager codes
\cite{mackay95} and its variations \cite{idosaad,VSK}.
 We believe that the connections found between belief networks and
statistical physics can be further developed to provide deeper insights into
the typical performance of general error-correcting codes.

\label{sec:conclusions}

%\acknowledgements
\ack
This work was partially supported by the program 
``Research For The Future'' (RFTF) of the Japanese Society for the 
Promotion of Science (YK) and by EPSRC grant GR/L52093 and a Royal Society travel grant (DS and RV).

%@@@@@@@@@@@@@@@@@@@@@@@@@@@@@@@@@@@@@@@@@@@@@@@@@@@@@@@@@@@@@@@@@@@@@@
\appendix

\section{Free Energy}
\label{app:A}

In order to compute  free-energies one needs to calculate the replicated 
partition function (\ref{eq:partit_2}). One can start from Eq. 
(\ref{eq:partit}):

\begin{equation}
\label{eq:partit_app0}
\fl \langle {\cal Z}^{n}\rangle_{{\cal A},\xi,J} = \mbox{Tr}_{\{S_j^\alpha\}}
\left[\left \langle \mbox{exp}\left(-\beta {\cal H}^{(n)}(\{\mbox 
{\boldmath $S^\alpha$  }\})
\right)\right\rangle_{{\cal A},J,\xi} \right],
\end{equation}
where ${\cal H}^{(n)}(\{\mbox{\boldmath$S^\alpha$}\})$ represents the replicated Hamiltonian and $\alpha$ the replica indices.
First one  averages over the parity check tensors $\cal A$, for that 
 an appropriate distribution has to be introduced, denoting $\mu\equiv
\langle i_1,...,i_K\rangle$ for a specific set of indices:
\begin{equation}
\label{eq:partit_app1}
\fl \langle {\cal Z}^{n}\rangle = \left \langle \frac{1}
{{\cal N}} \sum_{\{{\cal A}\}}\prod_{i}\delta\left(\sum_{\mu\setminus i} 
{\cal A}_{\mu} -C \right)\mbox{Tr}_{\{S_j^\alpha\}} \mbox{exp}
\left(-\beta \;{\cal H}^{(n)}(\{\mbox{\boldmath $S$}^\alpha\}) \right)\right\rangle_{J,\xi}, 
\end{equation}
where the  $\delta$ distribution imposes a restriction on the connectivity per spin, $\cal N$ is a normalization coefficient and the notation $\mu\setminus i$
means the set $\mu$ minus the element $i$.
Using integral representations for the delta functions and rearranging:
 \begin{equation}
\label{eq:partit_app2}
\fl \langle {\cal Z}^{n}\rangle =\mbox{Tr}_{\{S_j^\alpha\}} 
\left \langle \frac{1}{{\cal N}} \left(\prod_i \oint \frac{dz_i}{2\pi  i}
\frac{1}{z_i^{C+1}}\right)\sum_{\{{\cal A}\}}\left(\prod_{\mu}
(\prod_{i\in \mu}z_{i})^{{\cal A}_{\mu}}\right) \mbox{exp}\left(-\beta
{\cal H}^{(n)}(\{\mbox{\boldmath $S^\alpha$}\}) \right)\right\rangle_{J,\xi}. 
\end{equation}
Remembering that ${\cal A}\in\{0,1\}$, and using the expression 
(\ref{eq:Hamiltonian}) for the Hamiltonian one  can change  the order of the
summation and the product above and sum over ${\cal A}$:
\begin{eqnarray}
\label{eq:partit_app3}
\fl \langle {\cal Z}^{n}\rangle =\mbox{Tr}_{\{S_j^\alpha\}}
 \left \langle \frac{1}{{\cal N}} \left(\prod_i \oint \frac{dz_i}{2\pi  i}
 \frac{1}{z_i^{C+1}}\right)e^{\beta F\sum_{\alpha,i}
\xi_i S_i^{\alpha}}\right.\nonumber\\
\times \left. \prod_{\mu}
\left[1 +(\prod_{i\in\mu}z_{i}) \mbox{exp}\left(\beta J_{\mu}\sum_{\alpha}
\prod_{i\in\mu} S_{i}^{\alpha} \right)\right ]\right\rangle_{J,\xi}. 
\end{eqnarray}
Using the identity $\mbox{exp}(\beta J_{\mu} \prod_{i\in\mu}S_i^{\alpha})= 
\mbox {cosh}(\beta)\left[1+\left(\prod_{i\in\mu}S_i^{\alpha}\right)
\mbox{tanh}(\beta J_{\mu})\right]$ one  can perform the product over
$\alpha$ to write:

\begin{eqnarray}
\label{eq:partit_app3b}
\fl \langle {\cal Z}^{n}\rangle =\mbox{Tr}_{\{S_j^\alpha\}}
  \frac{1}{{\cal N}} \left(\prod_i \oint \frac{dz_i}{2\pi  i}
 \frac{1}{z_i^{C+1}}\right)\left \langle e^{\beta F\sum_{\alpha,i}
\xi_i S_i^{\alpha}}\right\rangle_{\xi}\\ \nonumber
\times\prod_{\mu}
\left[1 +\left(\prod_{i\in\mu}z_{i}\right) \mbox{cosh}^n(\beta)
\left(1+\langle \mbox{tanh}(\beta J) \rangle_{J}\sum_{\alpha}
\prod_{i\in \mu}S_{i}^{\alpha}\right.\right. \\ \nonumber
 + \left. \left.  \langle \mbox{tanh}^{2}(\beta J) \rangle_{J}
\sum_{\langle\alpha_1 \alpha_2\rangle}\prod_{i\in \mu}S_{i}^{\alpha_1} 
\prod_{j\in \mu} S_{j}^{\alpha_2}+ ... \right)\right]. 
\end{eqnarray}

Defining $\langle \mu_1,\mu_2,...,\mu_l \rangle$ as an ordered set
of sets, and observing that  for large $N$, $\sum_{\langle \mu_1 ...
\mu_l \rangle} (...) =\frac{1}{l!}\left(\sum_{\mu} (...) \right)^l$ one
can perform the product over the sets $\mu$ and replace the series that
appears by an exponential:
\begin{eqnarray}
\fl\langle {\cal Z}^{n} \rangle=\mbox{Tr}_{\{S_j^{\alpha}\}}
\frac{1}{{\cal N}}\left( \prod_i \oint \frac{dz_i}{2\pi i}
\frac{1}{z_i^{C+1}}\right)\left\langle e^{\beta F\sum{\alpha,i}\xi_i S_i^
{\alpha}}\right \rangle_{\xi}\\ \nonumber
\times \mbox{exp}\left[\mbox{cosh}^{n}(\beta) \left(
\sum_{\mu}(\prod_{i\in \mu} z_i)+\langle\mbox{tanh}(\beta J)\rangle_J
\sum_{\alpha}\sum_{\mu}\prod_{i\in\mu}z_i S_i^{\alpha} \right. \right. 
\\ \nonumber
+ \left. \left.\langle\mbox{tanh}^{2}(\beta J)\rangle_J 
\sum_{\langle \alpha_1 \alpha_2 \rangle}\sum_{\mu}
\prod_{i\in\mu}z_i S_i^{\alpha_1} S_i^{\alpha_2}+...\right)\right].
\end{eqnarray}

Observing that $\sum_{\mu}=1/K! \sum_{i_1,...i_K}$, defining 
${\cal T}_l=\langle \cosh^n(\beta J)\! \tanh^l(\beta J) \rangle_J$
and introducing  auxiliary variables
$q_{\alpha_1...\alpha_m}=\frac{1}{N}\sum_{i}z_i 
S_i^{\alpha_1}...S_i^{\alpha_m}$ one finds:

 \begin{eqnarray}
\label{eq:partit_app4}
\fl \langle {\cal Z}^{n}\rangle_{{\cal A},\xi,J} = 
 \frac{1}{{\cal N}}\left(\prod_i \oint \frac{dz_i}{2\pi  i} \frac{1}
 {z_i^{C+1}}\right)\left(\int\frac{ dq_0 d\widehat{q}_0}{2 \pi i} \right)
\left(\prod_{\alpha} \int\frac{dq_{\alpha} d\widehat{q}_{\alpha}}{2 \pi i}
\right) \ldots \\ \nonumber
\times \mbox{exp}\left[\frac{N^K}{K!} \left({\cal T}_0 q_0^K+{\cal T}_1
\sum_\alpha q_\alpha^K+{\cal T}_2\sum_{\langle \alpha_1 \alpha_2\rangle}
q_{\alpha_1 \alpha_2}^{K} + \dots \right)\right] \\ \nonumber
\times \mbox{exp}\left[-N \left(q_0\widehat{q}_0 + \sum_{\alpha}q_\alpha 
\widehat{q}_\alpha +\sum_{\langle \alpha_1 \alpha_2\rangle}q_{\alpha_1 
\alpha_2}\widehat{q}_{\alpha_1 \alpha_2}+\ldots  \right)\right]\\ \nonumber
\times \mbox{Tr}_{\{S_j^\alpha\}}\left[\left\langle e^{\beta F\sum_{\alpha,i}
\xi_i S_i^{\alpha}}\right\rangle_{\xi}\mbox{exp}\sum_i
\left(\widehat{q}_0z_i + \sum_{\alpha}\widehat{q}_\alpha  z_i S_i^\alpha
+\ldots\right)\right]. 
\end{eqnarray}

The normalization constant is given by:
\begin{equation}
\fl {\cal N}=\sum_{\{{\cal A}\}}\prod_{i} \delta\left( \sum_{\mu\setminus i}
{\cal A}_{\mu} - C \right),
\end{equation}
 and can be computed using exactly the same methods as above resulting in:
\begin{equation}
\fl {\cal N}=\left(\prod_i \oint \frac{dz_i}{2\pi i}\frac{1}{z_i^{C+1}}\right)
\left( \int \frac{dq_0 d\widehat{q}_0}{2\pi i}\right) \mbox{exp }\left[
 \frac{N^K}{K!} q_0^K - N q_0 \widehat{q}_0 + \widehat{q}_0 \sum_i z_i\right].
\end{equation} 

Computing the integrals over $z_i$'s and using  Laplace's method to 
compute the integrals over $q_0$ and $\widehat{q}_0$ one  gets:
\begin{equation}
\fl {\cal N}= \mbox{exp}\left\{\mbox{Extr}_{q_0,\widehat{q}_0}\left[
\frac{N^K}{K!}q_0^K -Nq_0 \widehat{q}_0 + N \mbox{ln}\left(
\frac{\widehat{q}_0^C}{C!} \right)\right] \right \}.
\end{equation}
The extremum point is given by $q_0=N^{(1-K)/K}[(K-1)!C]^{1/K}$ and 
$\widehat{q}_0=(C\,N)^{(K-1/K)}\left [(K-1)!\right]^{-1/K}$.
Replacing the Lagrange multipliers in Eq.(\ref{eq:partit_app4}) 
using $q_{\alpha_1...\alpha_m}/q_0 \rightarrow q_{\alpha_1...\alpha_m}$
and $\widehat{q}_{\alpha_1...\alpha_m}/q_0 \rightarrow 
\widehat{q}_{\alpha_1...\alpha_m}$, computing the integrals over $z_i$ and
using  Laplace's method to evaluate the integrals over the Lagrange
multipliers one finally find Eq.(\ref{eq:partit_2}).

\section{Replica Symmetric Solution}
\label{app:B}
The replica symmetric free-energy (\ref{eq:freesym}) can be obtained
by plugging the ansatz (\ref{eq:auxfields}) into Eq.(\ref{eq:partit_app4}).
After computing the normalization ${\cal N}$ and using  Laplace's
method one has:

\begin{equation}
\label{eq:RS1}
\fl\langle {\cal Z}^{n}\rangle_{{\cal A},\xi,J} =
\mbox{exp}\left\{N\,\mbox{Extr}_{\pi,\widehat{\pi}}
\left[ \frac{C}{K} {\cal G}_1\,-\,C\,{\cal G}_2 \,+\,{\cal G}_3 \right]
\right\},
\end{equation}
where:
\begin{eqnarray}
\fl {\cal G}_1={\cal T}_0 
+ {\cal T}_1 \sum_\alpha \int \prod_j^K \left(\,dx_j\,\pi(x_j)\,
\mbox{tanh}(\beta x_j)\right)\nonumber \\ 
+ {\cal T}_2\sum_{\langle \alpha_1 \alpha_2\rangle}
\int \prod_j^K \left(\,dx_j\,\pi(x_j)\,\mbox{tanh}^2(\beta x_j)
\right)  + \ldots, 
\end{eqnarray}

\begin{eqnarray}
\fl {\cal G}_2 = 1+ \sum_{\alpha} \int\,dx\,dy \,\pi(x)\,
\widehat{\pi}(y)\, \mbox{tanh}(\beta x)\, \mbox{tanh}(\beta y)
\nonumber\\
+\sum_{\langle \alpha_1 \alpha_2\rangle}\int\,dx\,dy \,\pi(x)\,
\widehat{\pi}(y)\, \mbox{tanh}^2(\beta x)\, \mbox{tanh}^2(\beta y)
+\ldots  
\end{eqnarray}
 and
\begin{eqnarray}
\fl {\cal G}_3=\frac{1}{N} \,\mbox{ln } \left \{
\left(\prod_i \oint \frac{dz_i}{2\pi  i} \frac{1} {z_i^{C+1}}\right)
\mbox{Tr}_{\{S_j^\alpha\}}\left[ \left\langle 
\mbox {exp }{\beta F\sum_{\alpha,i}\xi_i S_i^{\alpha}}
\right\rangle_{\xi}  \right. \right.  \nonumber \\ 
 \times \mbox{exp }\widehat{q}_0
\left(\sum_i\, z_i + \sum_{\alpha}\sum_i\, z_i S_i^\alpha \int \, dy \,
\widehat{\pi}(y) \mbox{tanh}(\beta y)\right. \nonumber \\
 + \left. \left. \left. \sum_{\langle \alpha_1 \alpha_2\rangle} 
\sum_i\, z_i S_i^{\alpha_1}S_i^{\alpha_2}
\int \, dy \, \widehat{\pi}(y) \mbox{tanh}^2(\beta y)
+\ldots \right)\right] \right\}. 
\end{eqnarray}

The equation for ${\cal G}_1$ can be worked out by using the definition of 
${\cal T}_m$ and
the fact that $(\sum_{\langle \alpha_1\ldots\alpha_l \rangle} 1)=$
 \small  $\left( \begin{array}{c} n \\ l \end{array} \right)$\normalsize
    to write:
\begin{equation}
\label{eq:g1series}
\fl {\cal G}_1=\left\langle\mbox{cosh}^n(\beta J) \int 
\left(\prod_{j=1}^K \,dx_j \, \pi(x_j)\right)
\left(1+\mbox{tanh}(\beta J)\prod_{j=1}^{K}
\mbox{tanh}(\beta x_j)\right)^n \right\rangle_J.
\end{equation}

Following exactly the same steps one  obtains:
\begin{equation}
\fl {\cal G}_2 = \int\,dx\,dy \pi(x)\,\widehat{\pi}(y)\, 
\left ( 1 + \mbox{tanh}(\beta x)\, \mbox{tanh}(\beta y)\right)^n,
\end{equation}
and
\begin{eqnarray}
\label{eq:g3series}
\fl {\cal G}_3 = \,\mbox{ln } \left \{
\mbox{Tr}_{\{S^\alpha\}}\left[ 
\left \langle \mbox {exp }\left({\beta F\xi\sum_{\alpha} S^{\alpha}}\right)
\right\rangle_{\xi}  \right. \right.  \nonumber \\ 
 \times \left. \left.
  \oint \frac{dz}{2\pi  i}\frac{1} {z^{C+1}}
 \mbox{exp }\left(\widehat{q}_0 \,z \int \, dy \, \widehat{\pi}(y)
 \prod_{\alpha=1}^{n}(1+S^\alpha \mbox{tanh}(\beta y))\right)
 \right] \right\}. 
\end{eqnarray}

Computing the integral over $z_i$ and the trace one finally finds:
\begin{equation}
\fl {\cal G}_3= \,\mbox{ln }\left \{ \frac{\widehat{q}_0}{C!}\int 
\prod_{l=1}^{C}\,dy_l \widehat{\pi}(y_l)
\left[\sum_{\sigma=\pm 1} \left\langle e^{\sigma \beta F \xi}\right\rangle_\xi \prod_{l=1}^C 
(1+\sigma\mbox{tanh}(\beta y_l)) \right]^n \right \}.
\end{equation}

Putting everything together, using Eq.(\ref{eq:freenergy}) and some simple 
manipulation one finds Eq.(\ref{eq:freesym}).

\section{Zero Temperature Self-consistent Equations}
\label{app:C}
In this appendix we describe how one can write a set of self-consistent equations to solve the zero temperature saddle-point equations (\ref{eq:sp_infty}). Supposing a three peaks ansatz given by:
\begin{eqnarray}
\widehat{\pi}(y)&=&p_+\delta(y-1)+p_0\delta(y)+p_-\delta(y+1)\\
\pi(x)&=&\sum_{l=1-C}^{C-1} T_{[p_{\pm},p_0;C-1]}(l)\; \delta(x-l), 
\end{eqnarray}
with
\begin{equation}
\fl T_{ \left[ p_{+}, p_{0}, p_{-}; C \right]} (l) = \sum_{\scriptsize \{k,h,m \ ; \
k-h=l \ ; \ k+h+m=C-1\}} \frac{(C-1)!}{k! \ h! \ m!} \ p_{+}^{k} \
p_{0}^{h} \ p_{-}^{m}.
\end{equation}
 One can consider the problem as a random walk, where
 $\widehat{\pi}(y)$ describes the probability of one step of length $y$ 
($y>0$ means one step to the right) and ${\pi}(x)$ describes the probability
 of being at distance $x$ from the origin after $C-1$ steps. With this idea 
in mind it is relatively easy to understand 
 $T_{ \left[ p_{+}, p_{0}, p_{-}; C-1 \right]} (l)$  as the probability 
of walking the distance $l$ after $C-1$ steps with the probabilities 
$p_{+}$, $p_{-}$ and  $p_{0}$  of  respectively moving right, left and staying
at the same position. We define the probabilities of walking right/left as  $\psi_{\pm} =\sum_l^{C-1} T_{ \left[ p_{+}, p_{0}, p_{-}; C-1 \right]} (\pm l)$.
Using  the second saddle-point equation (\ref{eq:sp_infty})  one can write:

\begin{equation}
\fl p_{+}=\int\left[\prod_{l=1}^{K-1} dx_l \, \pi(x_l) \right]
\left\langle \delta \left[1-\mbox{sign}(J\prod_{l=1}^{K-1}x_l) \mbox{min}
(\mid J\mid, \mid x_1 \mid, \ldots, \mid x_{K-1} \mid \right] \right\rangle_J
\end{equation}

The left side of the above equality can be read as the probability of 
making $K-1$ independent walks such that after $C-1$ steps 
 all of them are not in the origin  and an even (for$J=+1$) or odd (for $J=-1$) number of walks are at the left side. Using this reasoning for $p_{-}$  and
$p_{0}$  one can finally write :
\begin{eqnarray}
\fl p_{+}= (1-p)\sum_{j=0}^{\lfloor \frac{K-1}{2} \rfloor} 
\scriptsize \left(\begin{array}{c} K-1 \\ 2j \end{array} \right) \normalsize
\psi_{-}^{2j} \psi_{+}^{K-2j-1} + 
p\sum_{j=0}^{\lfloor \frac{K-1}{2} \rfloor - 1} 
 \scriptsize \left(\begin{array}{c} K-1 \\ 2j+1 \end{array} \right) \normalsize
\psi_{-}^{2j+1} \psi_{+}^{K-2j-2} \nonumber
\\ + p \,\psi_{-}^{K-1} \mbox{odd}(K-1) \\
\fl p_{-} =(1-p)\sum_{j=0}^{\lfloor \frac{K-1}{2} \rfloor - 1} 
\scriptsize \left(\begin{array}{c} K-1 \\ 2j+1 \end{array} \right) \normalsize
\psi_{-}^{2j+1} \psi_{+}^{K-2j-2} + 
p\sum_{j=0}^{\lfloor \frac{K-1}{2} \rfloor - 1} 
  \scriptsize\left(\begin{array}{c} K-1 \\ 2j \end{array} \right) \normalsize
\psi_{-}^{2j} \psi_{+}^{K-2j-1} \nonumber
\\ + (1-p)\, \psi_{-}^{K-1} \mbox{odd}(K-1),
\end{eqnarray}
where $\mbox{odd}(x)=1(0)$ if $x$ is odd (even).
Using that $p_{+}+p_{-}+p_{0}=1$ one can obtain $p_{0}$. A similar set of 
equations can be obtained for a five peaks ansatz leading to the same set
of solutions for the FERRO and PARA phases. The PARA solution $p_{0}=1$
is always a solution, for $C>K$ a FERRO solution with $p_{+}>p_{-}>0$ 
emerges.

\section{}
\label{app:nishifree}
In this appendix we establish 
the identity $\langle J \rangle_J = \langle J \mbox{ tanh}(\beta_N J)\rangle_J$  for symmetric channels. 
It was shown in \cite{sourlas94} that :
\begin{equation}
\fl \beta_N\,J=\frac{1}{2}\mbox{ln}\left(\frac{p(J\mid 1)}{p(J\mid -1)}\right),
\end{equation} 
where $\beta_N$ is the Nishimori's temperature and $p(J \mid J^0)$ are 
the probabilities that a transmitted bit $J^0$ is received as $J$.
From this we can easily find:
\begin{equation}
\fl \mbox{tanh }(\beta_N\,J)=\frac{p(J\mid 1)-p(J\mid -1)}{p(J\mid 1)+p(J\mid -1)}.
\end{equation}
In a symmetric channel ($p(J\mid -J^0)=p(-J\mid J^0)$), 
it is also represented as
\begin{equation}
\mbox{tanh }(\beta_N\,J)=\frac{p(J\mid 1)-p(-J\mid 1)}{p(J\mid 1)+p(-J\mid 1)}.
\end{equation}

Therefore,
\begin{eqnarray}
\fl \langle J\; \mbox{tanh }(\beta_N\,J)\rangle_J 
= 
\mbox{Tr}_J \;p(J\mid 1)\;\frac{J\;p(J\mid 1)}{p(J\mid 1)+p(-J\mid 1)}\nonumber\\
+\mbox{Tr}_J\; p(J\mid 1)\;
\frac{(-J)\;p(-J\mid 1)}{p(J\mid 1)+p(-J\mid 1)}\nonumber \\ 
\lo= 
\mbox{Tr}_J \;p(J\mid 1)\;\frac{J\;p(J\mid 1)}{p(J\mid 1)+p(-J\mid 1)}\nonumber\\
+\mbox{Tr}_J\; p(-J\mid 1)\;
\frac{J\;p(J\mid 1)}{p(-J\mid 1)+p(J\mid 1)}\nonumber\\ 
\lo=
\mbox{Tr}_J \;J\;p(J\mid 1) =\langle J \rangle_J.
\end{eqnarray}

\section*{References}
\bibliography{stat}

\end{document}